# Topological-phase effects and path-dependent interference in microwave structures with magnetic-dipolar-mode ferrite particles

M. Berezin, E.O. Kamenetskii, and R. Shavit

Microwave Magnetic Laboratory
Department of Electrical and Computer Engineering,
Ben Gurion University of the Negev, Beer Sheva, Israel

March 26, 2012

**Abstract**

Different ways exist in optics to realize photons carrying nonzero orbital angular momentum. Such photons with rotating wave fronts are called twisted photons. In microwaves, twisted fields can be produced based on small ferrite particles with magnetic-dipolar-mode (MDM) oscillations. Recent studies showed strong localization of the electric and magnetic energies of microwave fields by MDM ferrite disks. For electromagnetic waves irradiating MDM disks, these small ferrite samples appear as singular subwavelength regions with time and space symmetry breakings. The fields scattered by a MDM disk are characterized by topologically distinctive power-flow vortices and helicity structures. In this paper we analyze twisted states of microwave fields scattered by MDM ferrite disks. We show that in a structure of the fields scattered by MDM particles, one can clearly distinguish rotating topological-phase dislocations. Specific long-distance topological properties of the fields are exhibited clearly in the effects of path-dependent interference with two coupled MDM particles. Such double-twisted scattering is characterized by topologically originated split-resonance states. Our studies of topological-phase effects and path-dependent interference in microwave structures with MDM ferrite particles are based on numerical analysis and recently developed analytical models. We present preliminary experimental results aimed to support basic statements of our studies.

## I. INTRODUCTION

Present interest in electromagnetic fields with rotating wave fronts – twisted photons – stimulates investigations of such field structures in different frequency domains, from optical [1, 2] to radiowave frequencies [3, 4]. Recently, it has been shown that in microwaves, twisted fields can be produced based on small ferrite particles with magnetic-dipolar-mode (MDM) oscillations. The spatial distribution of the fields originated from a MDM particle is nonhomogeneous in the sense that the equal phase fronts are helices. In creation of these rotating-front microwave fields, a main role plays topological-phase effects [5 – 7].

The phase, known as the Berry phase, or the geometric phase, has found applications in a wide range of disciplines throughout physics. Geometric phases may show up whenever the system under study depends on several parameters and is transported around a closed path in parameters space [8]. While the derivation of the Berry phase makes use of the transport of states in Hilbert space, there are many classical derivations for geometric phases. Without having to appeal to quantum mechanics, one can consider the topological phases in such classical effects as, for example, noninertial motion due to parallel transport of coordinates [9] and polarization rotation in helicoidal fibers [10 – 13]. Due to the Berry phase one can observe a spin-orbit interaction in optics. It was shown that a spin-orbit interaction of photons results in fine splitting of levels in a ring dielectric resonator, similarly to that of electron levels in an atom [14]. Since the Berry phase is purely geometric in origin, it does not depend on dynamic considerations. It is also independent on the properties of the circuit in ordinary space.

Very specific properties of geometrical phases become evident in a case of rotational effects in magnetic samples. In Ref. [15] it was shown that sample rotation induces frequency splittings in nuclear-quadrupole-resonance spectra. The observed splittings can be interpreted in two ways: (a) as a manifestation of Berry's phase, associated with an adiabatically changing Hamiltonian, and (b) as a result of a fictitious magnetic field, associated with a rotating-frame transformation. The main standpoint is that the dynamical phase evolution in magnetic structure is unaffected by magnetic rotational motion and any rotational effects must arise from Berry's phase. Related effects were predicted in other magnetic resonance experiments that involve sample rotation. Among these experiments, it is worth noting ferromagnetic-resonance (FMR) rotating magnetic samples. Frequencies of the FMR are typically in the gigahertz range, which is far above achievable angular velocities of mechanical rotation of macroscopic magnets. In Refs. [16, 17] it was discussed that a rotating ferromagnetic nanoparticle (having a very small moment of inertia) can be a resonant receiver of the electromagnetic waves at the frequency of the FMR. If such a receiver is rotating mechanically at an angular velocity $\Omega$, one has splittings of the FMR frequency. The frequency of the wave perceived by the receiver equals

$$\omega^{(n)} = \omega_{FMR} - \Omega, \qquad (1)$$

where $\omega_{FMR}$ is frequency of the ferromagnetic resonance, $\Omega = \frac{n\hbar}{I}$, $I$ is the moment of inertia of a nanoparticle, and $n = 0, \pm 1, \pm 2, \pm 3, ...$ The experimental results for the rotational motion of solid ferromagnetic nanoparticles confined inside polymeric cavities give evidence for the quantization [16].

There exists, however, an evidence for the FMR splittings in macroscopic magnets without mechanical rotation of samples. The multiresonance splittings in the FMR spectra were observed experimentally in non-rotating macroscopic quasi-2D ferrite disks with magnetostatic-wave (MS-wave) [or magnetic-dipolar-mode (MDM)] oscillations [18 – 20]. In these microwave experiments, the MDM ferrite disk behaves as a multiresonant receiver for the electromagnetic waves irradiating this sample. Analytical and numerical studies of this effect gave a deep insight into an explanation of the experimental absorption spectra [5 – 7, 21 – 25]. It was shown that very effective multiresonance interactions of a ferrite disk with external electromagnetic fields are the result of energy eigenstates and specific topological properties of MDM oscillations. For electromagnetic waves irradiating a MDM ferrite disk, this small sample appears as a topological singularity characterized by strong 3D localization of electromagnetic energy with the power-flow-density vortices. Evidently, in such ferrite particles one has the Berry-phase generation not by the sample rotation, but due to topological characteristics of the fields of MDM oscillations.

When we discuss unique properties of the MDM oscillations in quasi-2D ferrite disks, an important question also arises: Could the observed effect of microwave energy localization due to small MDM ferrite particles (or, in other words, particles with the MS-magnon oscillations) be considered, somehow, as a dual effect of the energy localization in optics due to metallic particles with the electrostatic-plasmon oscillations? In a case of subwavelength metallic-sphere particles with plasmon oscillations, to answer the question: how can a particle absorb more than the light incident on it [26], one uses the classical Mie theory [27]. A small plasmon-oscillation particle localizes the electric energy of the electromagnetic wave. The scattered fields correspond to the fields scattered by a small electric dipole and the problem is one of solving the Laplace equation with the proper boundary conditions at a sphere surface [28]. An array of such particles can be treated as a chain of point electric dipoles [29, 30]. For small subwavelength ferrite-disk particles with MS-magnon oscillations, situation is completely different. In this case, one has strong localization of both the electric and magnetic energies in a subwavelength region of the electromagnetic wave. Moreover, in this subwavelength region, there is mutual coupling between the electric and magnetic fields resulting in appearance of the power-flow-density vortices [5, 7]. If one supposes that he has created a particle with a local (subwavelength-region) coupling between the electric and magnetic energies – a magnetoelectric (ME) particle – one, certainly, should demonstrate a special topological structure of the fields near this particle. Would the fields scattered by ME particles correspond to the fields scattered by a local combination of a small electric dipole and a small magnetic dipole? A positive answer to this question means that using a gedankenexperiment with two quasistatic, electric and magnetic, point probes for the near-field characterization, one should observe not only the electrostatic-potential distribution (because of the electric polarization) and not only the magnetostatic-potential distribution (because of the magnetic polarization), one also should observe a special cross-potential term (because of the cross-polarization effect). This fact contradicts to classical electrodynamics. What will be an expression for the Lorentz force acting between these particles? This expression should contain an "electric term", a "magnetic term", and a special "magnetoelectric term". Such an expression is unknown in classical electrodynamics. One cannot consider (in classical electrodynamics) a system of two coupled electric and magnetic dipoles as local sources of the fields and there are no two coupled Laplace equations (for the magnetostatic and electrostatic potentials) in the near-field region [28]. In a presupposition that a particle with an effect of local coupling between the electric and magnetic energies is really created, one has to show that inside this particle there are internal dynamical motion processes with special symmetry properties.

Theoretically, it was shown that dynamical processes of magnetic dipole-dipole interactions in quasi-2D ferrite disks are characterized by symmetry breaking effects. These effects are well modeled analytically with use of a so-called magnetostatic-potential (MS-potential) wave function $\psi$. The boundary-value-problem solutions for a scalar wave function $\psi$ in a quasi-2D ferrite disk demonstrate multiresonance MDM spectra with unique topological structures of the mode fields [5 – 7, 21 – 25]. Based on the $\psi$-function solutions, it was also shown that the near fields originated from ferrite-disk particles with MDM oscillations are characterized by topologically distinctive power-flow vortices, non-zero helicity, and a torsion degree of freedom [31, 32]. The existence of the power-flow-density vortices of the MDM oscillations presumes an angular momentum of the fields. The electric and magnetic fields of MDMs have both the "spin" and "orbital" angular momentums. One can find that in a source-free subwavelength region of microwave fields there exists the field structures with local coupling between the time-varying electric and magnetic fields differing from the electric-magnetic coupling in regular-propagating free-space electromagnetic (EM) waves. To distinguish such field structures, the term magnetoelectric (ME) fields has been used [31]. Small ferrite-disk particles with MDM spectra, being sources of microwave ME near fields, are termed as ME particles [31, 32]. The coupled states of electromagnetic fields with MDM vortices can be specified as the MDM-vortex polaritons [7].

The properties of the MDM oscillations inside a ferrite disk are well observed numerically due to the ANSOFT HFSS program. In a case of ferrite inclusions acting in the proximity of the FMR, the phase of the wave reflected from the ferrite boundary depends on the direction of the incident wave. For a given direction of a bias magnetic field, one can distinguish the right-hand and left-hand rays of electromagnetic waves. This fact, arising from peculiar boundary conditions of the fields on the dielectric-ferrite interfaces, leads to the time-reversal symmetry breaking effect in microwave resonators with inserted ferrite samples [33 – 36]. A nonreciprocal phase behavior on a surface of a ferrite disk results in path-dependent numerical integration of the problem. For the spectral-problem solutions, the HFSS program, in fact, composes the fields from interferences of multiple plane EM waves inside and outside a ferrite particle. One has a very good correspondence of the HFSS results with the analytical and experimental results. From the HFSS studies one can see the same eigenresonances and the same field topology of the resonance eigenstates in different microwave structures {when, for example, a ferrite disk is placed in a rectangular waveguide and when it is placed in a microstrip structure [32]}.

It becomes evident that the helicity properties, observed in the near-field (subwavelength) region of the MDM ferrite disk [5 – 7, 31, 32], can be transferred and transformed to the far-field (on scales compared with the incident EM wave) region. As we will show in this paper, a tiny ferrite disk changes the field topology in entire waveguide structure, having sizes much, much bigger than sizes of a ferrite particle. Based on numerical (the HFSS- program) results, we are able to observe topological structures of the fields scattered by the MDM ferrite particle and the long-distance helicity properties of the fields in a waveguide. We can see that on certain planes in vacuum, the fundamental excitations appear in loop-like forms. We show that in the field topology, there can be path-dependent loop quantization of the fields. We show also that on metallic waveguide walls, there exist electric and magnetic topological charges. We analyze the path-dependent interference in microwave waveguides due to two MDM particles placed at distances comparable with the disk diameters and at distances essentially exceeding sizes of these particles. The results show that because of the path-dependent integration, an entire volume of a waveguide is strongly involved in the process of an interaction between small-scale MDM particle oscillations and large-scale waveguide-mode fields. Our numerical studies are supported by recently developed analytical models for MDM oscillations in ferrite-disk particles. We present preliminary experimental results aimed to support some basic statements of our studies.

## II. MDM FERRITE DISKS AS AXION-LIKE PARTICLES

It has been shown [31, 32], that in a vacuum region, which is in close vicinity to a quasi-2D ferrite disk with the MDM oscillations, there exists a specific parameter of the field structure: the helicity density of the fields. For a real electric field, this parameter is written as

$$F \equiv \frac{1}{8\pi} \vec{E} \cdot \nabla \times \vec{E}. \tag{2}$$

For time-harmonic fields, the parameter is calculated as

$$F = \frac{1}{16\pi} \text{Im}\left\{ \vec{E} \cdot \left( \vec{\nabla} \times \vec{E} \right)^* \right\}. \tag{3}$$

One can also calculate a space angle between the vectors $\vec{E}$ and $\vec{\nabla} \times \vec{E}$ as

$$\cos\alpha = \frac{\text{Im}\left\{ \vec{E} \cdot \left( \vec{\nabla} \times \vec{E} \right)^* \right\}}{\left| \vec{E} \right| \left| \nabla \times \vec{E} \right|}. \tag{4}$$

This is the helicity density normalized to the field amplitudes. The presence of non-zero helicity parameter for the fields near small quasi-2D ferrite disks with the MDM oscillation spectra makes it evident the fact of existence of unique field structures – microwave ME fields. Such fields, originated from intrinsic dynamical process of magnetization inside a quasi-2D ferrite disk, have a topological-phase structure [5 – 7, 31]. In the vicinity of a ferrite disk, the fields are splitted for the regions with positive and negative helicity parameters [31].

The analytically derived topological characteristics of the MDM fields are well illustrated by the HFSS numerical results. In a case of numerical studies, one uses an entire *k*-space of electromagnetic waves. Ferrite inclusions are characterized by non-hard-wall boundary conditions with the interplay of reflection and transmission at the interfaces. These properties are combined with the time-reversal symmetry breaking effects. At the surface of a quasi-2D ferrite disk, the fields inside and outside a particle are related by the electromagnetic boundary conditions. At resonance frequencies, one has an enhanced effect of the field localization due to multiple back and forth bouncing of the fields inside a disk. It is very important that such a complicated process of numerical integration (when one has interferences of multiple plane EM waves inside and outside a ferrite particle) is well modeled analytically. In the analytical studies, however, an interaction of a small ferrite particle with MDM spectra is described by a quantum-like model, non-trivial from a classical point of view.

From the analytical consideration, observation of the helicity density in the field structure tells us about the presence of specific axion-like fields of MDM which interact with the external EM fields. The MDM ferrite disks appear as pseudo-scalar particles. The helicity density *F*, defined by Eq. (2), transforms as a pseudo-scalar under space reflection $\Pi$ and it is odd under time reversal T. With the observed helicity properties of the fields in a vacuum region, abutting to a quasi-2D ferrite disk with the MDM oscillations, one becomes faced with fundamental aspects akin to the problems of axion electrodynamics. Maxwell's equations are invariant when the electric and magnetic fields mix via rotation by an arbitrary angle $\xi$. For a real angle $\xi$, such a duality symmetry leaves invariant quadratic forms for the fields. This symmetry can be extended to Maxwell's equations in the presence of sources, provided that additional magnetic charges and currents are introduced [28]. Whenever a pseudo-scalar axion-like field $\vartheta$ is introduced in the theory, the dual symmetry is spontaneously and explicitly broken. An axion-electrodynamics term, added to the ordinary Maxwell Lagrangian [37]:

$$L_\vartheta = \kappa \vartheta \vec{E} \cdot \vec{B}, \qquad (5)$$

where $\kappa$ is a coupling constant, results in modified electrodynamics equations with the electric charge and current densities replaced by [37, 38]

$$\rho^{(e)} \to \rho^{(e)} + \kappa \vec{\nabla} \vartheta \cdot \vec{B} \quad \text{and} \quad \vec{j}^{(e)} \to \vec{j}^{(e)} - \kappa \left( \frac{\partial \vartheta}{\partial t} \vec{B} + \vec{\nabla} \vartheta \times \vec{E} \right). \qquad (6)$$

Integrating Eq. (5) over a closed space-time with periodic boundary conditions, we obtain the quantization

$$S_\vartheta = \int L_\vartheta d^4 x = \vartheta n, \qquad (7)$$

where *n* is an integer. It is evident that $S_\vartheta$ is a topological term. While $S_\vartheta$ generically breaks the parity and time-reversal symmetry, both symmetries are intact at $\vartheta = 0$ and $\vartheta = \pi$. The field $\vartheta$ itself is gauge dependent.

Axion-like fields and their interaction with the electromagnetic fields have been intensively studied and they have recently received attention due to their possible role played in building topological insulators. A topological isolator is characterized by additional parameter $\vartheta$ that couples the pseudo-scalar product of the electric and magnetic fields in the effective topological action. Evidently in common insulators $\vartheta = 0$. Interesting possibility assuming T-invariance exists when $\vartheta = \pi$, and this defines the new class of insulators – topological insulators [39 – 43]. The form of the effective action implies that an electric field can induce a magnetic polarization, whereas a magnetic field can induce an electric polarization. This effect is known as the topological ME effect and $\vartheta$ is related to the ME polarization. The topological angle $\vartheta$ is a static parameter for a topological insulator. Promoting $\vartheta$ to a dynamical field entails that the integrand in action (7) would change the field equation. In Ref. [42] it was pointed out that, when there is an antiferromagnetic order in an isolator, the magnetic fluctuations can couple to the electromagnetic fields, playing the role of the dynamical axion fields. In some materials, classified as topological magnetic isolators, the axion field couples linearly to light, resulting in the axionic polariton. One has a condensed-matter realization of the axion electrodynamics [42, 43].

In this paper we show that MDM ferrite disks can be considered as axion-like particles. The coupling of such pseudo-scalar particles with microwave photons plays a key role in our studies. A mechanism of interaction of electromagnetic fields with a small MDM ferrite disk is non-trivial. This is purely a topological effect. In topological isolators, the main concept of topological order appears due to the surface (or edge) states [39 – 41]. With a certain similarity to such surface states, we can see that in MDM particles, the topological effects are exhibited due to chiral edge modes on a lateral surface of a ferrite disk [23]. This becomes evident from the spectral properties of MDMs in such a particle. To make the MDM spectral problem analytically integrable, two approaches were suggested. These approaches, distinguishing by differential operators and boundary conditions used for solving the spectral problem, give two types of the MDM oscillation spectra in a quasi-2D ferrite disk. We, conventionally, name these two approaches as the *G*- and *L*-modes in the magnetic-dipolar spectra [6, 25]. The MS-potential wave function $\psi$ manifests itself in different manners for every of these types of spectra. For *L* modes, the MS-potential wave functions $\psi$ are odd under both parity Π and time-reversal T, and are ΠT -invariant [7]. The *L*-mode solutions show existence of four helical MS waves. Specific interactions of these waves give evidence for the double-helix resonances. In a case of the *L* modes, the MS-potential wave function $\psi$ is considered as a generating function for the vector harmonics of the magnetic fields [5 – 7, 24, 25]. The eigen electric and magnetic fields of the MDM oscillations one derives from the *L*-mode solutions. The helicity density for the fields, shown in Eqs. (3), (4), is obtained based on the *L*-mode spectrum [6, 7, 31]. In a case of the *G*-mode spectrum, where the physically observable quantities are energy eigenstates and eigen electric moments of a MDM ferrite disk, the MS-potential wave function $\psi$ appears as a Hilbert-space scalar wave function. There is a univocal correspondence between the *G*-mode and *L*-mode spectrum solutions. As it was shown in Ref. [7], there exists a certain operator $\hat{C}$ which transforms the *L*- mode spectrum to the energy eigenstate spectrum of MS-mode oscillations – the *G*-modes. For *G*-modes, we have magnetically gapped chiral edge states. These edge states concern the concept of topological order. Topological phases of the edge states are accumulated by the double-valued border functions on a lateral surface of a ferrite disk and the topological effects become apparent through the integral fluxes of the pseudo-electric fields. There are positive and negative fluxes corresponding to the counterclockwise and clockwise edge-function chiral rotations.

To have the standard action as a Lorentz scalar [28], one should multiply the helicity parameter of MDM by a certain scalar function, which is odd under both Π and T. Such a scalar function appears due to topological-phase edge states on a lateral surface of a ferrite disk. Coupling of MDM disks to external EM fields arises exclusively due to interaction of these fields with the eigen electric fluxes and eigen electric (anapole) moments originated from the MDM topological-phase edge states [23]. The following below studies give evidence for such ME coupling of the axion-like MDM oscillations to microwave EM fields.

**III. SCATTERING OF ELECTROMAGNETIC WAVES BY SMALL MDM PARTICLES**

We consider scattering of electromagnetic fields by MDM ferrite disks in a structure a rectangular waveguide with an enclosed ferrite particle. For better illustration of our studies of scattering effects by small ferrite particles, we use S-band rectangular waveguide which has the sizes of a waveguide cross section (4 cm × 10 cm) much bigger than sizes of a ferrite disk (the thickness of 0.05 mm and the diameter of 3 mm).

A normally magnetized ferrite disk is placed inside a waveguide (operating at the $TE_{10}$ mode) symmetrically with respect to waveguide walls so that the disk axis is perpendicular to a wide wall of a waveguide. Fig. 1 shows geometry of the structure. For a numerical analysis in the present paper, we use the yttrium iron garnet (YIG) disk normally magnetized by a bias magnetic field $H_0 = 2,930 \, \text{Oe}$; the saturation magnetization of the ferrite is $4\pi M_s = 1880 \, \text{G}$. For better understanding the field structures we use a ferrite disk with very small losses: the linewidth of a ferrite is $\Delta H = 0.1 \, \text{Oe}$. The numerically calculated reflection coefficient, represented in Fig. 2, clearly verifies the known spectral properties of MDMs [5, 7]. For the first two modes, the peak positions of the numerically obtained spectrum in Fig. 2 are in good correlation with the MDM spectrum calculated based on an analytical model [24, 25]. However, because of a very big difference between sizes of a waveguide and a ferrite resonator, one cannot observe such a good correlation for higher-order MDMs.

Our analysis in the present paper we restrict with consideration of topological phase effects only for the 1st MDM. The unique helicity properties of the fields in the vicinity of a MDM ferrite disk become evident from Fig. 3. In this figure one can see the normalized helicity density calculated numerically based on Eq. (4) at the resonance frequency of the 1st MDM ($f$ = 3.0237GHz). The pictures are obtained in two mutually perpendicular cross-section planes passing through the disk axis. The main point of the pictures shown in Fig. 3 is the fact that due to a MDM ferrite disk, a symmetrical Maxwellian-field structure in vacuum becomes splitted into two non-symmetrical (with non-zero helicities) field structures. These splittings are extended on extremely large area. One can see that the region of non-zero helicity parameters may cover almost an entire waveguide cross section. Along a waveguide axis, an entire region of non-zero helicity stretches about a quarter of the waveguide wavelength. On metal walls of a waveguide, the helicity is zero. We can say that below and above a ferrite disk the fields are twisted in different azimuth directions, being "pinned" at metal. This gives evidence for the torsion degree of freedom of the fields coupled to a MDM ferrite disk [31, 32].

The topological-phase effects in an entire microwave-waveguide structure with an enclosed MDM ferrite disk are well illustrated by the pictures of the time and space transformations of the wave fronts of the electric field. For the 1st resonance frequency of 3.0237GHz, Fig. 4 shows transformations of the wave front of the electric field in a waveguide on a vacuum plane 0.1 mm above a ferrite disk at different time-phases. The inserts show enlarged pictures of the electric-field distributions immediately above a ferrite disk. One can see that as the time-phase changes, the wave fronts change their geometrical forms. It is very important to note that a geometrical structure of the shown fronts restores twice during a time period of the microwave radiation. For two next-coming fronts of incident waveguide-mode fields with the time-phase difference of $\pi$, one can see the same configuration of the wave fronts distorted due to scattering from a MDM particle. The wave front becomes distorted, when passing through a peculiar topological relief. This relief, appearing due to non-zero field helicity, one has only at frequencies of the MDM resonances, when the helicity parameter is non-zero. The pictures of the $\vec{E}$-field structures at the non-resonance frequency and at the MDM-resonance frequency, shown in Fig. 5, allows compare configurations of the undistorted wave front and the wave front distorted due to a topological relief with non-zero helicity.

Together with strong localization of the electric field at the MDM resonances, one observes also strong localization of the magnetic field at the resonance frequencies. Fig. 6 shows the magnetic-field distributions on a vacuum plane 0.1 mm above a ferrite disk at different time phases. The frequency is the 1$^{st}$ resonance frequency of 3.0237GHz. The top row shows enlarged pictures of the magnetic-field distributions immediately above a ferrite disk. In the bottom row one can see structures of the wave fronts of the magnetic field. A character of the magnetic field near a MDM ferrite disk looks like the field of a rotating magnetic dipole. Evidently, for magnetic fields, there are no transformations of geometrical forms of wave fronts, as we can observe for electric fields. Strong localization of both the electric and magnetic fields near MDM particles results in appearance of power-flow vortices at resonance frequencies. For the 1$^{st}$ resonance frequency of 3.0237GHz, such a vortex on a vacuum plane 0.1 mm above a ferrite disk is shown in Fig. 7. Such topological structures of the power-flow density of the near fields at the MDM resonances were studied in details in Refs. [6, 7, 24].

The field distributions near a small ferrite disk, shown in Figs. 3 – 7, give evidence for the fact that the observed scattering phenomena are very different from the known classical effects of scattering of electromagnetic fields by small particles [27, 44]. The field in vacuum, surrounding a ferrite-disk particle, is a superposition of the incident (the waveguide-mode) and scattered (the MDM) fields:

$$\vec{H}' = \vec{H}_{WG} + \vec{H}_{MDM}, \qquad (9)$$

$$\vec{E}' = \vec{E}_{WG} + \vec{E}_{MDM}. \qquad (10)$$

The incident waveguide-mode EM field excites MDM egenstates in a small quasi-2D ferrite disk. There is a very strong difference between wavenumbers of the waveguide-mode EM wave and the MDM oscillations. We have $k_{MDM} \gg k_{EM}$. Based on the field structures shown in Figs. 3 – 7 (and also based on our previous studies [6, 7]), we can state that interaction of electromagnetic fields with a small MDM ferrite particle cannot be considered as scattering by small electric or/and magnetic dipoles. In particular, one cannot find any similarity between scattering by a MDM ferrite particle and scattering by a small electric-dipolar particle with plasmon oscillations [26 – 30].

How do incident EM fields interact with a MDM ferrite particle? From the pictures in Fig. 6 one can see that in the region close to a ferrite disk, the magnetic field $\vec{H}_{MDM}$ appears as the field of a magnetic dipole rotating in the *xy* plane. A small ferrite disk is placed in a waveguide symmetrically with respect o waveguide walls. On this place, the field $\vec{H}_{WG}$ is linearly polarized with respect to the *x* axis (see Fig. 1). With representation of this field as two circular polarized fields, rotating in the *xy* plane, one can clearly see that magnetic-dipolar field $\vec{H}_{MDM}$ is perpendicular to both circular polarized magnetic fields of a waveguide mode. So no magnetic interaction of a MDM particle with an incident magnetic field occurs. Does the particle interact with the electric component of an incident EM field and, if does, what is the nature of this interaction? One of the main features of the shown above field distributions is the fact that in the near-field vacuum region immediately above a ferrite disk there are only in-plane components of the field $\vec{E}'$. The field $\vec{E}_{MDM}$ is perpendicular to the field $\vec{E}_{WG}$ and, evidently, a normal component of the field $\vec{E}'$ is zero at every time phase $\omega t$ during a time period of microwave oscillations. It means that in the near-field vacuum region immediately above a ferrite disk, the scattered electric field completely annihilates the incident electric field. Since the incident field $\vec{E}_{WG}$ is oriented perpendicularly to the disk plane, this is an evidence for the property of a MDM ferrite disk to create an eigen electric-field flux in opposition to the flux of the externally applied RF electric field. Spectral properties of the MDMs show that on a lateral surface of a quasi-2D ferrite disk there exist the double-valued-function loop magnetic currents which produce eigen electric fields and hence eigen electric fluxes through the loop. In this case one can introduce a notion of an electric self inductance as the ratio of the electric flux to the magnetic current. In a region far from a MDM disk, it can be described as a small particle with eigen electric moment counter-directed to the field $\vec{E}_{WG}$. Such an eigen electric moment – the anapole moment – was described in Refs. [23, 31, 45, 46]. The anapole moment of a ferrite disk arises not from the curl electric fields of MDM oscillations. An analysis of symmetry properties of the anapole moment allows understanding the mechanism of interaction of MDM particle with an external electric field. In Refs. [20, 45, 46], it was shown experimentally that the MDM ferrite disks can effectively interact with the external RF electric field oriented normally to the disk plane.

The observed interaction of an external EM field with a MDM ferrite disk becomes understandable when we consider more in detail the topological properties of MDMs. Analytically, the helicity density of the MDM near fields is derived based on the properties of the *L*-mode spectra [6, 31]. At the same time, the energy content (weight) of each MDM is determined by the *G*-modes. The egenstates of the operator $\hat{C}$ (which transforms the *L*- mode spectrum to the *G*-mode energy eigenstate spectrum) are chiral edge states on a lateral surface of a ferrite disk. The edge geometrical-phase loops are penetrated by the fluxes of the pseudo-electric field [7, 23]. The edge geometrical-phase loops are created by two helices. This arises from a geometrical structure of *L* modes, which are helical MS modes [6]. In the near-field region, the electric-field distribution can be well illustrated based on an analysis of the helical topological loops composed with branches of two (left-handed and right-handed) helices. Fig. 8 shows transformations of the electric fields in a region very close to a ferrite disk. The electric-field vectors are shown in connection with geometry of the MS helical-wave loops [6]. During the dynamical-phase variation of $\pi$, one has a complete circulation along a double-helix topological loop, but with the electric-field vectors directed oppositely.

In the region far from a ferrite disk, one has a more complicated situation. A more detailed analysis of the electric-field structure clearly shows a topological character of the scattered fields. It becomes evident that the topologically originated scattered electric fields have components in the *xy* plane and are oppositely directed with respect to *z* axis. For a given time phase $\omega t$, Fig. 9 (*a*), (*b*) show the electric field distributions in vacuum above and below a ferrite disk, respectively, at the distances 0.1mm from the ferrite-disk planes and at a bias magnetic field directed along *z* axis. During the dynamical-phase variation, the topological loops are transformed both in the form and in orientation. Moreover, the topological loops can be partially closed by metallic walls of a waveguide. This is illustrated in Fig. 9 (*c*), (*d*). There are the electric field distributions at the same vacuum planes as in Fig. 9 (*a*), (*b*), but at another time phase $\omega t$. Configurations of the geometrical-phase loops are recreated for every $\pi$ phase difference of the field vectors. The regions of non-zero helicity are the regions of pseudo-electric-field fluxes. When a bias magnetic field is directed opposite to the *z*-axis direction, one has different orientations of both a wave front and electric-field vectors. This is illustrated in Fig. 9 (*e*), where the time phase and the vacuum plane position (0.1 mm above a ferrite disk plane) are the same as in Fig 9 (*d*). In an entire space inside a waveguide, we have the geometrical-phase shells for the electric-field vectors. The topological-phase loops in Fig. 9 are observed on the *xy* cross-sectional planes of these shells. Fig. 10 shows the topological-phase loops for other cross-sectional planes: the *xz* and *yz* planes inside a waveguide. The pictures in Fig. 10 give evidence for transformations of the electric fields in accordance with a model of the MS helical-wave loops (see Fig. 8). It is worth noting that the topological-phase profiles are correlated with profiles of the helicity parameter. In Fig. 11, we show the electric-field-front configuration on a vacuum plane 7 mm above a ferrite-disk plane combined with the cross-sectional outline of the helicity density on the same plane.

The structure of the field scattered by a small ferrite particle with MDM oscillations has properties resembling the field structure of axion-electrodynamics. We have non-zero helicity density $F$, transformed as a pseudo-scalar under space reflection Π with odd time reversal T [31, 32], and a pseudo-scalar axion-like field originated from the topological-order chiral edge modes on a lateral surface of a ferrite disk [23, 31]. If the axion fields exist, they will couple the radiation EM field to the MDM field with a coupling proportional to the helicity density. The fact that the MDM oscillations in a ferrite disk, coupled to the electromagnetic fields, can play the role of the dynamical axion fields is also well illustrated by distributions of the currents and charges induced on metal walls of a waveguide. Fig. 12 shows modifications of the surface electric charge and current densities due to the MDM-oscillation particle. Presumably, these modifications can be described by the axion-electrodynamics terms in Eq. (6). In Fig. 12, we can see that surface electric charges, appearing also as topological charges, originate surface electric currents in the forms of convergent or divergent spirals. Because of such a topological effect, an entire structure of the current distributions on the waveguide walls is strongly distorted. As unique feature of the observed topological effect, we can point out on the presence of topological magnetic charges on the waveguide walls. This is illustrated in Fig. 13 for the upper wide wall of a waveguide. Because of symmetry of the magnetic fields in a waveguide along $z$-axis, one has the same pictures of topological magnetic charges on the lower waveguide wall. It is worth noting also that the regions of topological electric charges completely coincide with the regions of topological magnetic charges. The superimposed positions of these topological charges are shown in Fig. 14.

For better understanding of the shown numerical results (and, especially, for further analysis of interactions between two MDM ferrite disks) we have to dwell more in details on the answer to the above question: How do the incident single-valued EM fields interact with double-valued fields of MDM ferrite particles? The $G$-mode solutions for magnetic-dipolar oscillations are characterized by the single-valued MS-potential membrane functions. However, to satisfy the boundary conditions for magnetic flux density on a lateral surface of a disk, one has to impose a geometrical phase factor. This phase factor appears due to double-valued edge wave functions [23, 31]. For a $G$-mode membrane MS-wave function $\tilde{\eta}$, the boundary condition on a lateral surface of a ferrite disk is the following:

$$\mu\left(\frac{\partial \tilde{\eta}}{\partial r}\right)_{r=\Re^-} - \left(\frac{\partial \tilde{\eta}}{\partial r}\right)_{r=\Re^+} = 0, \qquad (11)$$

where $\Re$ is a radius of the disk. On a lateral border of a ferrite disk, the correspondence between a double-valued membrane wave function $\tilde{\varphi}$ and a singlevalued function $\tilde{\eta}$ is expressed as: $(\tilde{\varphi}_\pm)_{r=\Re^-} = \delta_\pm (\tilde{\eta})_{r=\Re^-}$, where $\delta_\pm \equiv f_\pm e^{-iq_\pm \theta}$ is a double-valued edge wave function on contour $L = 2\pi\Re$. The azimuth number $q_\pm$ is equal to $\pm\frac{1}{2}l$, where $l$ is an odd quantity ($l$ = 1, 3, 5, …). For amplitudes we have $f_+ = -f_-$ and $|f_\pm| = 1$. Function $\delta_\pm$ changes its sign when $\theta$ is rotated by $2\pi$ so that $e^{-iq_\pm 2\pi} = -1$. As a result, one has the energy-eigenstate spectrum of MS-mode oscillations with topological phases accumulated by the edge wave function $\delta$. On a lateral surface of a quasi-2D ferrite disk, one can distinguish two different functions $\delta_\pm$. A line integral around a singular contour $L$: $\frac{1}{\Re}\oint_L \left(i\frac{\partial \delta_\pm}{\partial \theta}\right)(\delta_\pm)^* dL = \int_0^{2\pi} \left[\left(i\frac{\partial \delta_\pm}{\partial \theta}\right)(\delta_\pm)^*\right]_{r=\Re} d\theta$ is an observable quantity. It follows from the fact that because of such a quantity one can restore singlevaluedness (and, therefore, Hermicity) of the $G$-mode spectral problem. Because of a geometrical phase factor on a lateral boundary of a ferrite disk, $G$-modes are characterized by a pseudo-electric field (the gauge field) [23, 31]. We denote this pseudo-electric field by the letter $\vec{\epsilon}$. The geometrical phase factor in the $G$-mode solution is not single-valued under continuation around a contour $L$ and can be correlated with a certain vector potential $\vec{\Lambda}_\epsilon^{(m)}$ [23, 31]:

$$i\Re \int_0^{2\pi} [(\vec{\nabla}_\theta \delta_\pm)(\delta_\pm)^*]_{r=\Re} d\theta \equiv K \oint_L \left(\vec{\Lambda}_\epsilon^{(m)}\right)_\pm \cdot d\vec{L} = 2\pi q_\pm. \tag{12}$$

where $\vec{\nabla}_\theta \delta_\pm = \frac{1}{\Re}\frac{\partial \delta_\pm}{\partial \theta}\bigg|_{r=\Re} \vec{e}_\theta$ and $\vec{e}_\theta$ is a unit vector along an azimuth coordinate, and $K$ is a normalization coefficient [31]. In Eq. (12) we inserted a connection which is an analogue of the Berry phase. In our case, the Berry's phase is generated from the broken dynamical symmetry. The confinement effect for magnetic-dipolar oscillations requires proper phase relationships to guarantee single-valuedness of the wave functions. To compensate for sign ambiguities and thus to make wave functions single valued we added a vector-potential-type term $\vec{\Lambda}_\epsilon^{(m)}$ (the Berry connection) to the MS-potential Hamiltonian. On a singular contour $L = 2\pi\Re$, the vector potential $\vec{\Lambda}_\epsilon^{(m)}$ is related to double-valued functions. It can be observable only via the circulation integral over contour $L$, not pointwise. The field $\vec{\epsilon}$ is the Berry curvature. In contrast to the Berry connection $\vec{\Lambda}_\epsilon^{(m)}$, which is physical only after integrating around a closed path, the Berry curvature $\vec{\epsilon}$ is a gauge-invariant local manifestation of the geometric properties of the MS-potential wavefunctions. The corresponding flux of the gauge field $\vec{\epsilon}$ through a circle of radius $\Re$ is obtained as:

$$K\int_S \left(\vec{\epsilon}\right)_\pm \cdot d\vec{S} = K\oint_L \left(\vec{\Lambda}_\epsilon^{(m)}\right)_\pm \cdot d\vec{L} = K\left(\Xi^{(e)}\right)_\pm = 2\pi q_\pm, \tag{13}$$

where $\left(\Xi^{(e)}\right)_\pm$ are quantized fluxes of pseudo-electric fields. There are the positive and negative eigenfluxes. These different-sign fluxes should be inequivalent to avoid the cancellation. It is evident that while integration of the Berry curvature over the regular-coordinate angle $\theta$ is quantized in units of $2\pi$, integration over the spin-coordinate angle $\theta'$ $\left(\theta' = \frac{1}{2}\theta\right)$ is quantized in units of $\pi$. The physical meaning of coefficient $K$ in Eqs. (11), (13) concerns the property of a flux of a pseudo-electric field. It is related to the notion of a magnetic current in the $G$-mode analysis. In Refs. [23, 45, 46], the coefficient $K$ was conventionally taken as equal to unit. In our recent study [31], we related quantity $K$ to an elementary flux of the pseudo-electric field. In the next section we show that quantity *K* can be found in an analysis of two interacting MDM ferrite disks.

The velocities of two border flows $\vec{\nabla}_\theta \delta_\pm$ results in two edge magnetic currents on a lateral surface of a ferrite disk [6, 23, 31]. For the observer placed in a laboratory frame, there will be two double-valued-function magnetic currents rotating at the same direction of the azimuth coordinate. In the vicinity of a ferrite disk, the linear-polarized $\vec{E}$-field of the incident wave can be decomposed in two (clockwise and counterclockwise) circularly rotating fields. On the other hand, for a given direction of a DC magnetic field $\vec{H}_0$, one has two azimuth modes of pseudo-electric fields originated from double-valued-function edge magnetic currents and rotating at a certain direction on a lateral surface of a ferrite disk. These azimuth modes of pseudo-electric fields are $\pi$-shifted. Fig. 15 illustrates the mechanism of interaction between the flux of the incident single-valued electric field and the flux of the double-valued pseudo-electric fields of MDM ferrite particles. From these pictures, one can see that (with averaging at the azimuth coordinates) the electric flux of the incident EM field is completely annihilated by the electric flux of the MDM field.

## IV. PATH-DEPENDENCE INTERFERENCE WITH TWO COUPLED MDM PARTICLES

We consider now interactions between two MDM ferrite disks placed in a microwave waveguide. When the spin-coordinate and orbital degrees of freedom of MDM oscillations are mixed in a particular way, the momenta of these oscillations in a coupled-particle system feel important effects of this $\pi$ Berry's phase. The shown above topological-effect distributions of the currents and charges induced on metal walls of a waveguide by a single MDM ferrite disk presume very specific conditions on coupling between these particles. When the MDM of one ferrite disk is rotating, it swirls its surrounding spacetime around. Rotating spacetime can impart to the field of a waveguide mode an intrinsic form of orbital angular momentum. This intrinsic form of orbital angular momentum can be detected by another ferrite disk. Interaction between two MDM ferrite disks can be non-reciprocal.

In Ref. [47], two interacting MDM ferrite disks were studied analytically based on the coupled-mode theory for *G*-mode spectra and with taking into account the pseudo-electric fluxes. It was shown that for such a structure, there can be two states of MS interaction with the symmetrical and antisymmetrical field distributions. It was also shown that the MDM interaction due to pseudo-electric fluxes appears with certain quantization rules for these fluxes. In both types of interactions (MS interaction and pseudo-electric-flux interaction), the coupled-state resonances should be distinguished by energy. There are two energetically stable states. To a certain extent, we have a model which rather more resembles coupled semiconductor quantum dots [48 – 52], than coupled plasmon-oscillation particles [53 – 55]. In this paper we present the results of numerical studies of two interacting MDM ferrite disks. These numerical results illustrate, sufficiently well, the theoretical model in Ref. [47]. In our numerical analysis, we distinguish two cases. In the first case (the short-distance interaction, when a gap between ferrite disks is much less than or comparable with diameters of disks) both the MS and pseudo-electric-flux interactions should be taken into account. In the second case (the long-distance interaction, when a gap is much bigger than diameters of disks) we have the coupling mainly due to pseudoelectric fluxes.

The MS interaction between two MDM disks can be analyzed based on the coupled-mode theory [47]. Interaction by the pseudo-electric fluxes for two laterally coupled MDM disks appears due to formation of closed topological loops between the disks. In an assumption about ability of the edge function in one location to produce phase accumulation in the edge function in another location, the electric interaction presumes existence of four pseudo-electric fluxes [47]. For a given MDM $p$, we will designate a pseudo-electric flux which is connected with an edge function of a ferrite disk $a$ and penetrating the border loop of disk $a$, as $\left(\Xi^e\right)_p^{aa}$, and a pseudo-electric flux which is connected with an edge function of a ferrite disk $b$ and penetrating the border loop of disk $b$, as $\left(\Xi^e\right)_p^{bb}$. At the same time, we will designate a pseudo-electric flux connected with an edge function of a ferrite disk $b$ and penetrating the border loop of disk $a$, as $\left(\Xi^e\right)_p^{ab}$, and a pseudo-electric flux connected with an edge function of a ferrite disk $a$ and penetrating the border loop of disk $b$, as $\left(\Xi^e\right)_p^{ba}$. To preserve the single-valuedness of the membrane wave function $\tilde{\eta}$ of disks $a$ and $b$ we have [47]

$$\left(\Xi^e\right)_p^{aa} + \left(\Xi^e\right)_p^{ab} = \left(\Xi^e\right)_p^{a} = \frac{1}{K} 2\pi q' \tag{14}$$

and

$$\left(\Xi^e\right)_p^{bb} + \left(\Xi^e\right)_p^{ba} = \left(\Xi^e\right)_p^{b} = \frac{1}{K} 2\pi q''. \tag{15}$$

Both $q'$ and $q''$ take values of $\pm\frac{1}{2}, \pm\frac{3}{2}, \pm\frac{5}{2}, ...$

Because of linearity, the quantities $\left(\Xi^e\right)_p^{ab}$ and $\left(\Xi^e\right)_p^{ba}$ can be represented as

$$\left(\Xi^e\right)_p^{ab} = k_p^{ab} \left(\Xi^e\right)_p^{b} = \frac{1}{K} 2 k_p^{ab} \pi q' \tag{16}$$

and

$$\left(\Xi^e\right)_p^{ba} = k_p^{ba} \left(\Xi^e\right)_p^{a} = \frac{1}{K} 2 k_p^{ba} \pi q''. \tag{17}$$

For mode $p$, ME-interaction coefficients $k_p^{ab}$ and $k_p^{ba}$ determine, respectively, a fraction of a total pseudo-electric flux of disk $a$ perceiving the border ring of disk $b$ and a fraction of a total pseudo-electric flux of disk $b$ perceiving the border ring of disk $a$. Evidently, $0 \leq k_p^{ab} \leq 1$ and $0 \leq k_p^{ba} \leq 1$. Based on Eqs. (14) – (17), we obtain

$$\left(\Xi^e\right)_p^{aa} = \frac{1}{K} 2\pi q' \left(1 - k_p^{ab}\right) \tag{18}$$

and

$$\left(\Xi^e\right)_p^{bb} = \frac{1}{K} 2\pi q'' \left(1 - k_p^{ba}\right). \tag{19}$$

Because of the path-dependence interaction, coupling between the disks is non-reciprocal. It means that $k_p^{ab} \neq k_p^{ba}$. So $\left(\Xi^e\right)_p^{aa} \neq \left(\Xi^e\right)_p^{bb}$ and $\left(\Xi^e\right)_p^{ab} \neq \left(\Xi^e\right)_p^{ba}$.

From Eqs. (14), (15), we have

$$\left(\Xi^e\right)_p^a - \left(\Xi^e\right)_p^b = \frac{1}{K} 2\pi \left(q' - q''\right). \tag{20}$$

The minimal difference between quantities $q'$ and $q''$ is $\pm 1$. This allows representing the coefficient $K$ as

$$K = \frac{2\pi}{\left|\left(\Xi^e\right)_p^a - \left(\Xi^e\right)_p^b\right|_{\min}}. \tag{21}$$

The fluxes $\left(\Xi^e\right)_p^a$ and $\left(\Xi^e\right)_p^b$ are normalized quantities with respect to the disk sizes and amplitudes of the MDM oscillations [31, 47]. Since a minimal difference of the pseudo-electric fluxes, $\left|\left(\Xi^e\right)_p^a - \left(\Xi^e\right)_p^b\right|_{\min}$, is a constant quantity, for a given geometry of MDM disks one should observe the quantum-like effect of the topological coupling between the disks. One can presume that because of two energetically stable states in the coupled-disk structure, the frequency difference between the resonance peak positions should not strongly depend on the distance between the disks. However, the intensity of the fields in the disks should depend on this distance. Moreover, since the coupling between the disks is non-reciprocal, there should be observable difference between the field intensities in the disks. Our numerical studies well illustrate these statements.

A structure of two coupled MDM disks embedded in a rectangular waveguide is shown in Fig. 16. For the short-distance interaction (the distance between centers of the disks is 3.6 mm), one observes splitting of the resonance peak. Fig. 17 shows this splitting on the reflection-coefficient characteristic of the 1st MDM. The frequency difference between the splitted peaks is very small. It is equal to 3.9 MHz. Such frequency difference is about 0.1% of the frequency of the 1st MDM. For the short distances, the main coupling due to MS interaction is accompanied with the auxiliary effect of the pseudo-electric-flux interaction. In Fig. 18 one can see the electric field distributions on the upper surface of a ferrite disks at a certain time phase. One has the symmetrical and antisymmetrical coupled modes. For both modes, there is an observable difference between the field intensities in the disks. In a case of the symmetrical mode, the field intensity for the disk *a* is higher than for the disk *b*, while for the antisymmetrical mode, one observes opposite situation with the field intensity. We can conclude that for the symmetrical mode, the flow of the pseudo-electric field piercing the disk *a* exceeds the flow piercing the disk *b*. For the antisymmetrical mode, one has the opposite situation. Fig. 19 shows positions of the wave fronts (localized regions where the electric field of a waveguide mode changes its sign) on a vacuum plane passing 0.1 mm above ferrite disks. While in a case of the symmetrical mode two disks interact with the incident EM field like one disk (compare with Fig. 4), for the antisymmetrical mode two disks interact with the incident EM field with a specific manner. In the last case, one has a very complicated structure of topological loops. It is worth noting that for two (symmetrical and antisymmetrical) types of modes, both disks are involved in the process of interaction with the incident EM field. Moreover, similarly to the situation with a single ferrite disk, a normal component of the field $\vec{E}'$ above every disk is zero at any time phase $\omega t$.

Let us consider now a structure when two disks are separated at the distance (between centers of the disks) of 6 mm. In this case, the splitting on the reflection-coefficient characteristic of the 1st MDM shown in Fig. 20 corresponds to the frequency of 3.8 MHz. For such a distance between the disks, the main coupling is due to the pseudo-electric-flux interaction. In Fig. 21 one can see the electric field distributions on the upper surface of a ferrite disks, corresponding to the symmetrical and antisymmetrical coupled modes. For these modes, one observes a strong difference between the field intensities in the disks. For the coupled-mode resonances, Fig. 22 shows positions of the wave fronts on a vacuum plane passing 0.1 mm above ferrite disks. In comparison with the case of the short-distance interaction, now only one disk is mainly involved in the process of interaction with the incident EM field. There is disk *a* for the symmetrical-mode resonance and disk *b* for the antisymmetrical-mode resonance. The helicity-density distributions for the coupled-mode resonances, shown in Fig. 23, give more evidence for the non-reciprocal interaction between the disks. One can see that for identical disks, the behaviors of coupling between the disks $a \leftarrow b$ and $a \rightarrow b$ are completely different. These topological effects take place only at the coupled-mode resonance. Fig. 24 shows the wave fronts on a vacuum plane 0.1 mm above ferrite disks for the frequency 3.0333GHz, which is between the coupled-mode resonances. There are no topological distortions of the wave fronts. It is also possible to show that at this frequency, the helicity density is zero.

As a very attractive feature of the topological-effect coupling, there is evidence for long-distance interaction between two MDM disks. In Fig. 25, we present the split-resonance characteristic of the 1st MDM at the distance between centers of the disks of 12 mm. The frequency split is about 1 MHz. Our studies confirm the fact that there is possibility to observe the topological-effect coupling between the disks at extremely long distances between the disks. Fig 26 shows a structure of two MDW ferrite disks placed inside a waveguide at the distance of 30 cm. The split-resonance characteristic of the 1st MDM at this distance is shown in Fig. 27. The frequency split is about 1.6 MHz. At the resonance frequencies, one has a strong difference between the field intensities in the disks. This is illustrated in Fig. 28, where one can see the helicity-density distributions for the coupled-mode resonances.

**V. EXPERIMENTAL SUDIES ON INTERACTION BETWEEN TWO MDM DISKS**

As a final stage of our studies on the path-dependent interference, we present some preliminary experimental results on interaction between two MDM disks. There are supplementary results aimed only to support some of our basic conclusions of this paper. More detailed experimental studies are the purpose of our future publications. For better sensitivity, in our experiment we use a microstrip structure with embedded ferrite disks. As we discussed in Introduction, for the same parameters of ferrite disks, the resonance frequencies of MDM oscillations should be the same in different microwave structures. Fig. 29 shows a microstrip structure with one ferrite disk and a numerical spectrum for this structure. Comparison of the numerical spectra in Figs. 2 (*a*) and 29 (*b*) shows that for different types of the external EM fields (fields of a waveguide and fields of a microstrip structure) one observes the same peak positions for the main eigenmodes of MDM oscillations.

For experimental studies, we used a microstrip structure with two disks. This structure is shown in Fig. 30 (*a*). The disks have diameter of 3 mm. They are made of the yttrium iron garnet (YIG) film on the gadolinium gallium garnet (GGG) substrate. The YIG film thickness is 50 mkm. The saturation magnetization of YIG is $4\pi M_0 = 1880 \text{G}$ and the linewidth is $\Delta H = 0.8 \text{Oe}$. It is necessary to note here that instead of a bias magnetic field in numerical studies ($H_0 = 2,930 \text{Oe}$), in the experiments we applied lower quantity of a bias magnetic field: $H_0 = 2,607 \text{Oe}$. Use of such a lower quantity (giving us the same positions of the resonance peaks in the numerical study and in the experiment) is necessary because of non-homogeneity of an internal DC magnetic field in a real ferrite disk. A more detailed discussion on a role of non-homogeneity of an internal DC magnetic field in the MDM spectral characteristics can be found in Refs. [19, 22].

Fig. 30 (*b*) shows the numerical and experimental normalized spectra for a microstrip structure with two coupled disks. The distance between the centers of the disks is 9 mm. One can observe the split-resonance characteristics for the 1$^{st}$ MDM. It is worth noting that in a microstrip structure, the frequency splits at the coupled-state responses are different compared to such responses in a rectangular waveguide. These splits (7.2 MHz for the numerical characteristic and 8.9 MHz for the experimental result) are more than 0.2% of the frequency of an incident EM field. As it follows from Fig. 30 (*b*), there is a sufficiently good correspondence between the numerical and experimental results for a microstrip structure with two coupled disks. This gives a preliminary confirmation of our numerical studies and analytical models. It is necessary to note, however, that the experimental peaks in Fig. 30 (*b*) are not so sharp and not identical in comparison with the peaks obtained from numerical results. The reasons for this are the following. We have different linewidths for ferrite disks in the numerical ($\Delta H = 0.1\,\text{Oe}$) and experimental ($\Delta H = 0.8\,\text{Oe}$) studies. Also, in experiments, we were unable to use absolutely identical disks: there are some differences with the disk geometries and material parameters.

## VI. DISCUSSIONS

A non-integrable (path-dependent) electromagnetic problem of ferrite disks placed inside a rectangular waveguide, following from closed-loop nonreciprocal phase behaviors on lateral surfaces of the disks, can be solved numerically based on the HFSS program. One of the main questions for discussions concerns credibility of the results, obtained based on the classical-electrodynamics HFSS program, for description of the shown non-trivial topological effects originated from the MDM ferrite disks. Also, a more serious question arises: Whether these numerical results are applicable for confirmation of such analytically-derived notions as pseudo-electric-field fluxes, anapole moments, energy eigenstates, etc., which hardly can be classified as pure classical notions?

First of all, we have to note strong regularity of the results obtained based on the HFSS-program solutions. From numerical studies of different microwave structures with embedded thin-film ferrite disks, one can see consistent pattern of both the spectral-peak positions and specific topological structure of the fields of the oscillating modes. Moreover, these numerical results are in very good correlations with the corresponding results obtained from analytical studies and microwave experiments. The HFSS program, in fact, composes the field structures from interferences of multiple plane EM waves inside and outside a ferrite particle. In such a numerical analysis one obtains the pictures of the field structures (resulting in the real-space integration) based on integration in the *k*-space of the EM fields. In the *k*-space integration, the fields are expanded from very low wavenumbers (free-space EM waves) to very high wavenumbers (the region of MS oscillations). As a result of numerical integration, one has convergent solutions at a given frequency. It is worth noting, however, that inside a small ferrite particle (and in the near-field region outside this particle), such a very complicated EM-wave process in a numerical representation is well modeled analytically with use of the MS-potential wave function $\psi$. As we discussed before, to make the path-dependent spectral problem of MDM oscillations analytically integrable, two approaches – the *G*- and *L*-mode magnetic-dipolar spectra – have to be considered. Evidently, inside (and nearly outside) a ferrite disk, the HFSS numerical results can be well applicable for confirmation of the analytically-derived quantum-like models based on scalar wave function $\psi$. There exist many examples in physics when non-classical effects are well modeled by a combined effect of numerous classical sub-elements.

As a very interesting effect of the topological-effect coupling, there is observation of the long-distance interaction between two MDM disks. In a waveguide structure with two interacting disks, the frequency split in the resonance characteristics is extremely small. Its width is about 0.1% of the frequency of an incident waveguide field. As we can see from our numerical simulation, the wave front of the fields scattered by the MDM ferrite disks can be oriented at a small angle (sometimes parallel) to the waveguide axis. It means the presence of the extremely-long-wavelength field structures in a waveguide. Such field structure may presume the quasi-static interaction between the particles. But how this can be realized at distances, which are very long in comparison with sizes of interacting particles? These distances are not small enough that retardation effects can be ignored. It can be supposed that there should be a certain mechanism of transportation of energy. As we showed above, there exist topological sources which appear on waveguide walls near the places of the location of ferrite disks. Could these topological sources cause a flow of topological-vortex (geometrical-phase) fluctuations against the background of a laminar EM-wave flow? The classical-type model cannot comprehensively explain the observed topological effects. Together with the above question, the open questions are: Why the frequency split in the resonance characteristics is almost independent on the distance between disks? Why the intensities of the fields in the disks are different for different frequencies of the split-resonance state. Based on our analytical semi-classical model, some answers to these questions can be suggested. The far-field interaction, can presume the closed-loop topology of pseudo-electric fluxes for the coupled-disk structure, which can be observed at very long distances inside a waveguide. In this case, the eigen and mutual pseudoelectric fluxes can be or summed or subtracted when they pierce a separate ferrite disk. When the fluxes are summed, the intensity of the fields near a disk will be high. Contrarily, when the fluxes are subtracted, the intensity of the fields near a disk will be low. We may follow the above mechanism of an interaction with the external EM field. Near every (of the coupled) particle the interaction of an external EM field is with an integrated flux of the pseudoelectric field. For two different interactions, one has two separate eigen frequencies. This is a quantized effect, which does not depend on distances between the disks. The eigen electric fluxes are quantized. The incident EM field cannot "see" this quantization for a single particle, but can recognize this quantization for coupled particles at the long-distance interaction. Further (and more detailed) studies on these unique topological effects are needed.

**VII. CONCLUSION**

Interaction of MDM particles with microwave radiation has the topological-effect nature. The presence of helicity density of the fields originated from MDM ferrite particles gives evidence for axion-like properties of these fields. Because of these properties, topology of the twisted states of microwave fields scattered by MDM ferrite disks strongly distinguish from a structure of an optical twisted photon.

In this paper, we have investigated the topological phase effects in a microwave waveguide with embedded MDM ferrite disks. We have shown that at MDM resonances, structures of the fields engineered by ferrite-disk particles (both in the near- and far-field regions) are substantially different from the properties of electromagnetic fields originated from usual microwave sources. Topological structures of the fields give evidence for the regions of phase singularities. One can see spontaneous generation of curved and closed-loop phase dislocations. The observed continuous variations of topological-contour pictures give evidence for a certain effect of the topological-wave propagation. In a system of coupled disks one has topologically originated split-resonance states. Specific topological properties of the fields of theses split-resonance states appear due to path-dependent interference of two coupled MDM particles. We have found that the split-resonance response for coupled MDM particles is weakly dependent on distances between disks. Such a response one can observe at distances, which are very long in comparison with sizes of interacting particles. These distances are not so small that one can ignore the electromagnetic-wave retardation effects. For proper physical explanations of the shown topological-phase effects and path-dependent interference in microwave structures with MDM ferrite particles, we suggested some analytical models. We presented preliminary experimental results on interaction between two MDM disks, which give confirmation of our numerical studies and analytical models.

**Figure captions**

Fig. 1. Geometry of a structure: a normally magnetized ferrite disk inside a rectangular waveguide. For better illustration of our studies of scattering effects by small ferrite particles, we use S-band rectangular waveguide which has the sizes of a waveguide cross section (4 cm × 10 cm) much bigger than sizes of a ferrite disk (the thickness of 0.05 mm and the diameter of 3 mm).

Fig. 2. MDM resonances of a quasi-2D ferrite disk. (*a*) Numerically obtained reflection coefficient. (*b*) Analytically derived spectral peak positions for the first three MDM resonances.

Fig. 3. The normalized helicity density, calculated based on Eq. (4), at the resonance frequency of the 1$^{st}$ MDM ($f$ = 3.0237GHz). (*a*) Cross-section plane *yz* (orthogonal to a front of an incident EM wave), passing through the disk axis; (*b*) cross-section plane *xy* (parallel to a front of an incident EM wave), passing through the disk axis.

Fig. 4. Transformations of the wave front of the electric field in a waveguide (at the 1$^{st}$ resonance frequency of 3.0237GHz) on a vacuum plane 0.1 mm above a ferrite disk at different time-phases. The inserts show enlarged pictures of the electric-field distributions immediately above a ferrite disk.

Fig. 5. The pictures of the $\vec{E}$-field structures on a vacuum plane 0.1 mm above a ferrite disk. (*a*) Undistorted wave front at frequency 3.1363GHz, which is between first and second MDM resonances; (*b*) distorted wave front at the 1$^{st}$ resonance frequency of 3.0237GHz.

Fig. 6. Magnetic-field distributions on a vacuum plane 0.1 mm above a ferrite disk at different time phases. The frequency is the 1$^{st}$ resonance frequency of 3.0237GHz. The top row: enlarged

pictures of the magnetic-field distributions immediately above a ferrite disk. The bottom row: structures of the wave fronts of the magnetic field.

Fig. 7. Power-flow vortex on a vacuum plane 0.1 mm above a ferrite disk. The frequency is the 1st resonance frequency of 3.0237GHz.

Fig. 8. Transformations of the electric fields in a region very close to a ferrite disk. The electric-field vectors are shown in connection with geometry of the MS helical-wave loops [20].

Fig. 9. The topologically originated scattered electric fields with the *xy*-plane components. (*a*), (*b*). The electric field distributions in vacuum above and below a ferrite disk, respectively, at the distances 0.1mm from the ferrite-disk planes and at a bias magnetic field directed along *z* axis. The time phase is $\omega t = 6°$. (*c*), (*d*). The electric field distributions in vacuum above and below a ferrite disk, respectively, at the distances 0.1mm from the ferrite-disk planes and at a bias magnetic field directed along *z* axis. The time phase is $\omega t = 15°$. (*e*) The same as in Fig. 9(*d*), but at a bias magnetic field directed opposite to the *z*-axis direction.

Fig. 10. The electric-field distrubution on the cross sectional planes at the 1st resonance frequency. (*a*) The *xz* plane in a waveguide; (*b*) The *yz* in a waveguide.

Fig. 11. The electric-field-front configuration on a vacuum plane 7 mm above a ferrite-disk plane combined with the cross-sectional outline of the helicity density on the same plane.

Fig. 12. Modification of the electric charge and current densities on waveguide walls at the 1st resonance frequency. Surface electric charges, appearing also as topological charges, originate surface electric currents in the forms of convergent or divergent spirals. (*a*) The electric charge and current densities on upper waveguide wall; (*b*) enlarged pictures of distributions on upper waveguide wall. (*c*) The electric charge and current densities on lower waveguide wall; (*d*) an enlarged picture of distributions on lower waveguide wall.

Fig. 13. Modification of the magnetic field on upper waveguide wall at the 1st resonance frequency. There is evidence for topological magnetic charges on the waveguide walls. Because of symmetry of the magnetic fields in a waveguide along *z*-axis, one has the same pictures of topological magnetic charges on the lower waveguide wall. (*a*) The magnetic field on upper waveguide wall; (*b*) enlarged pictures of distributions on upper waveguide wall.

Fig. 14. A combined picture of the magnetic-field (arrows) and surface current (lines) distributions on upper waveguide wall at the 1st resonance frequency. (*a*) Enlarged area (with the image of a ferrite disk) for the phase $\omega t = 21°$; (*b*) the same, for the phase $\omega t = 30°$.

Fig. 15. The mechanism of interaction between the flux of the incident single-valued electric field and the flux of the double-valued pseudo-electric fields of MDM ferrite particles. With averaging at the azimuth coordinates, the electric flux of the incident EM field is completely annihilated by the electric flux of the MDM field. (*a*), (*b*) Two (clockwise and counterclockwise) circularly rotating $\vec{E}$-fields of the incident wave. (*c*), (*d*) Two azimuth modes of pseudo-electric fields originated from double-valued-function edge magnetic currents and rotating at a certain direction on a lateral surface of a ferrite disk.

Fig. 16. A structure of two coupled MDM disks embedded in a rectangular waveguide.

Fig. 17. Reflection coefficient for two coupled disks at the frequency region of the 1st MDM. The distance between centers of the disks is 3.6 mm.

Fig. 18. The electric field distribution on the upper surface of ferrite disks. The distance between centers of the coupled disks is 3.6 mm. (*a*) At frequency (3.0253GHz) of the first peak of the splitted resonance; (*b*) at frequency (3.0292GHz) of the second peak of the splitted resonance.

Fig. 19. Positions of the wave fronts (localized regions where the electric field of a waveguide mode changes its sign) for coupled disks on a vacuum plane passing 0.1 mm above ferrite disks. The distance between centers of the disks is 3.6 mm. (*a*) At frequency (3.0253GHz) of the first peak of the splitted resonance; (*b*) enlarged pictures of the same distributions. (*c*) At frequency (3.0292GHz) of the second peak of the splitted resonance; (*b*) enlarged pictures of the same distributions.

Fig. 20. Reflection coefficient for two coupled disks at the frequency region of the 1st MDM. The distance between centers of the disks is 6 mm.

Fig. 21. The electric field distribution on the upper surface of ferrite disks. The distance between centers of the coupled disks is 6 mm. (*a*) At frequency (3.031GHz) of the first peak of the splitted resonance; (*b*) at frequency (3.0348GHz) of the second peak of the splitted resonance.

Fig. 22. Positions of the wave fronts (localized regions where the electric field of a waveguide mode changes its sign) for coupled disks on a vacuum plane passing 0.1 mm above ferrite disks. The distance between centers of the disks is 6 mm. (*a*) At frequency (3.031GHz) of the first peak of the splitted resonance; (*b*) enlarged pictures of the same distributions. (*c*) At frequency (3.0348GHz) of the second peak of the splitted resonance; (*b*) enlarged pictures of the same distributions.

Fig. 23. The normalized helicity density for coupled disks at cross-section plane *yz* (orthogonal to a front of an incident EM wave), passing through the disk axes. The distance between centers of the disks is 6 mm. (*a*) At frequency (3.031GHz) of the first peak of the splitted resonance; (*b*) at frequency (3.0348GHz) of the second peak of the splitted resonance.

Fig. 24. The wave front of the electric field for two coupled disks at frequency 3.0333GHz between two MDM resonances. Vacuum plane is 0.1 mm above ferrite disks. The distance between centers of the disks is 6 mm.

Fig. 25. Reflection coefficient for two coupled disks at the frequency region of the 1st MDM. The distance between centers of the disks is 12 mm.

Fig. 26. A structure of two far-distant MDM disks embedded in a rectangular waveguide. The distance between the disks is 30cm.

Fig. 27. Reflection coefficient for two coupled disks at the frequency region of the 1st MDM. The distance between the disks is 30cm.

Fig. 28. The normalized helicity density for coupled disks at cross-section plane *yz* (orthogonal to a front of an incident EM wave), passing through the disk axes. The distance between the disks is 30cm. The upper picture: frequency (3.0353GHz) of the first peak of the splitted

resonance. The lower picture: frequency (3.0369GHz) of the second peak of the splitted resonance.

Fig. 29. A microstrip structure with an embedded ferrite disk. (*a*) Geometry of a structure (a bias magnetic field is directed along *z* axis); (*b*) a numerical spectrum for the structure (a transmission coefficient between ports 1 and 2).

Fig. 30. A microstrip structure with two coupled ferrite disks. (*a*) Geometry of a structure (a bias magnetic field is directed along *z* axis, the distance between disks is 9mm); (*b*) numerical and experimental split-resonance characteristics for the 1$^{st}$ MDM (a normalized transmission coefficient between ports 1 and 2).

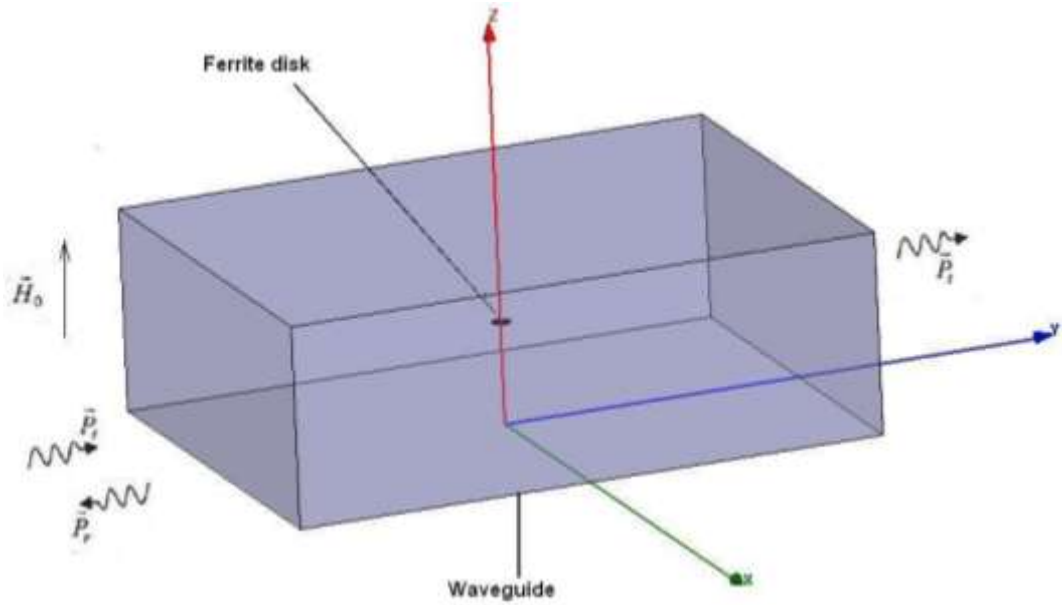

Fig. 1. Geometry of a structure: a normally magnetized ferrite disk inside a rectangular waveguide. For better illustration of our studies of scattering effects by small ferrite particles, we use S-band rectangular waveguide which has the sizes of a waveguide cross section (4 cm × 10 cm) much bigger than sizes of a ferrite disk (the thickness of 0.05 mm and the diameter of 3 mm).

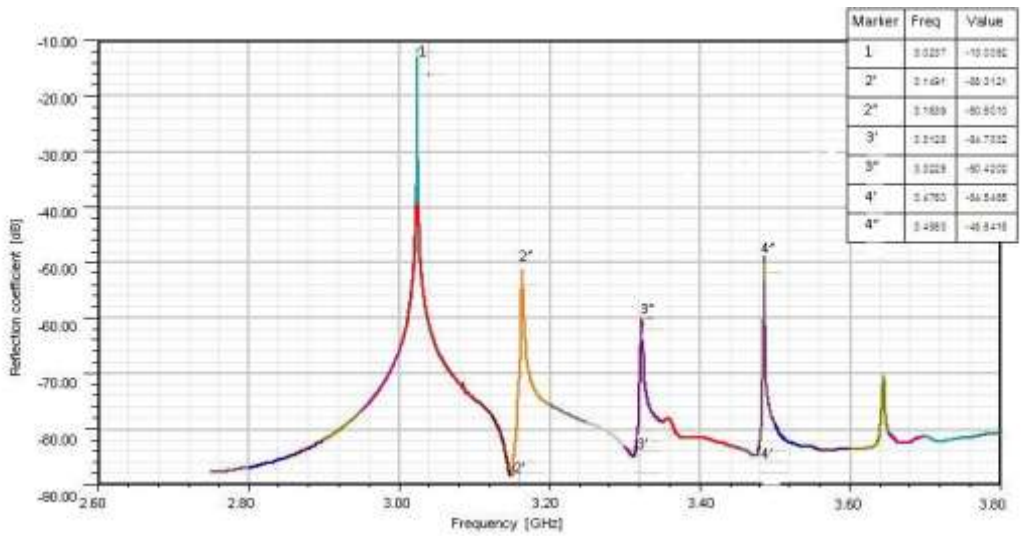

(a)

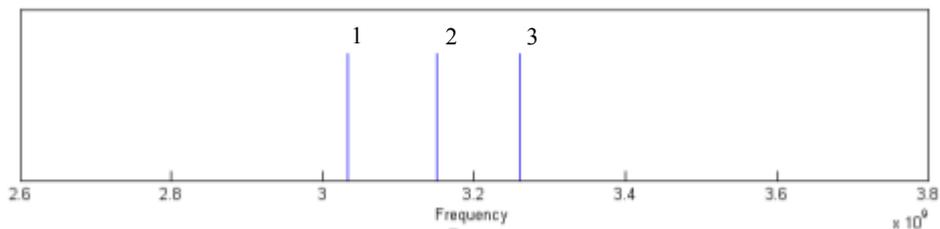

(b)

Fig. 2. MDM resonances of a quasi-2D ferrite disk. (*a*) Numerically obtained reflection coefficient. (*b*) Analytically derived spectral peak positions for the first three MDM resonances.

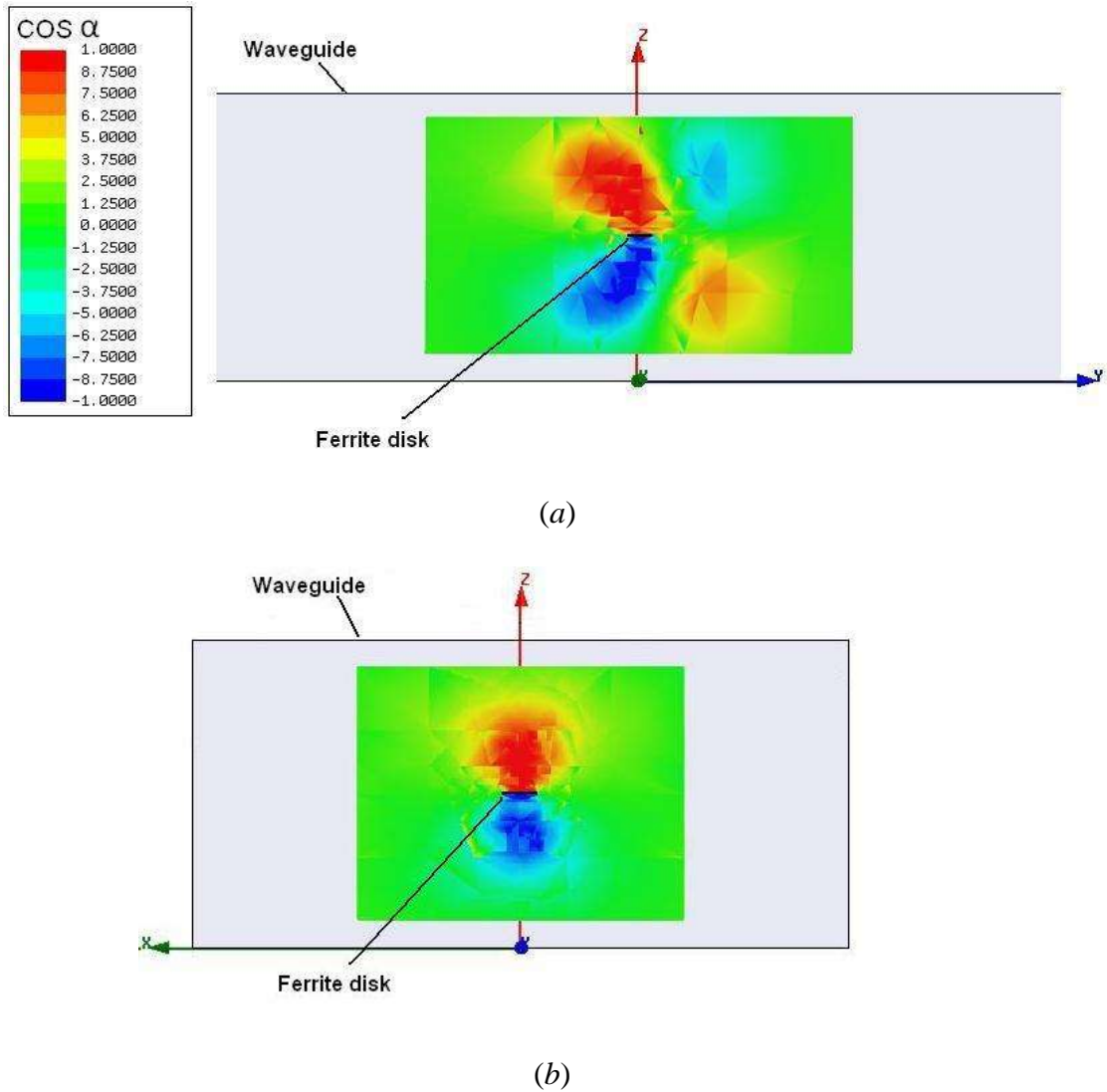

(*a*)

(*b*)

Fig. 3. The normalized helicity density, calculated based on Eq. (4), at the resonance frequency of the 1[st] MDM (*f* = 3.0237GHz). (*a*) Cross-section plane *yz* (orthogonal to a front of an incident EM wave), passing through the disk axis; (*b*) cross-section plane *xy* (parallel to a front of an incident EM wave), passing through the disk axis.

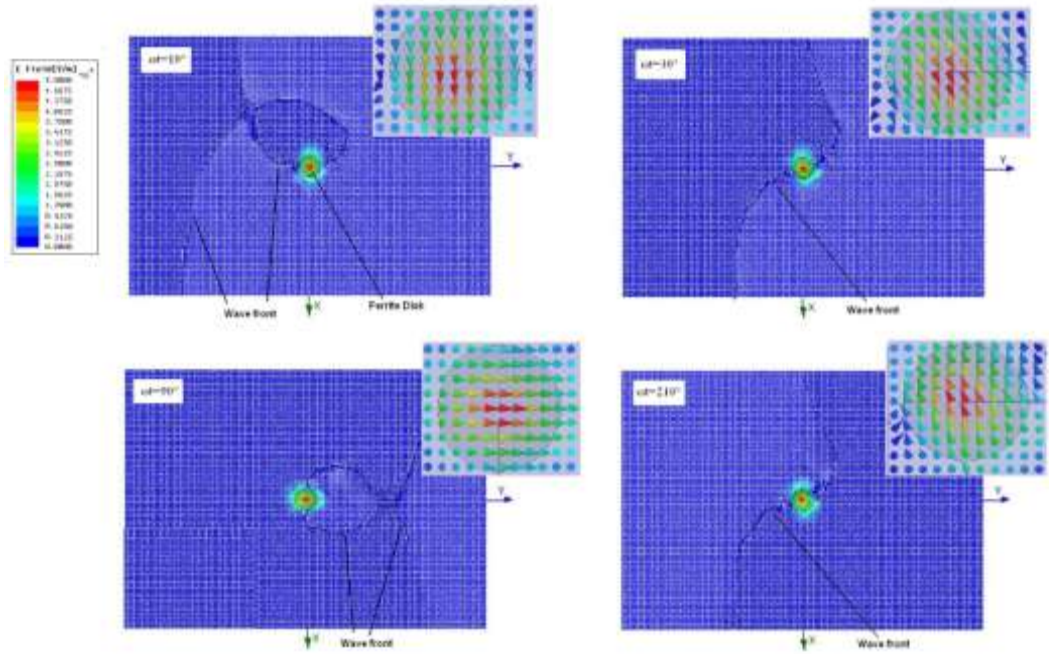

Fig. 4. Transformations of the wave front of the electric field in a waveguide (at the 1st resonance frequency of 3.0237GHz) on a vacuum plane 0.1 mm above a ferrite disk at different time-phases. The inserts show enlarged pictures of the electric-field distributions immediately above a ferrite disk.

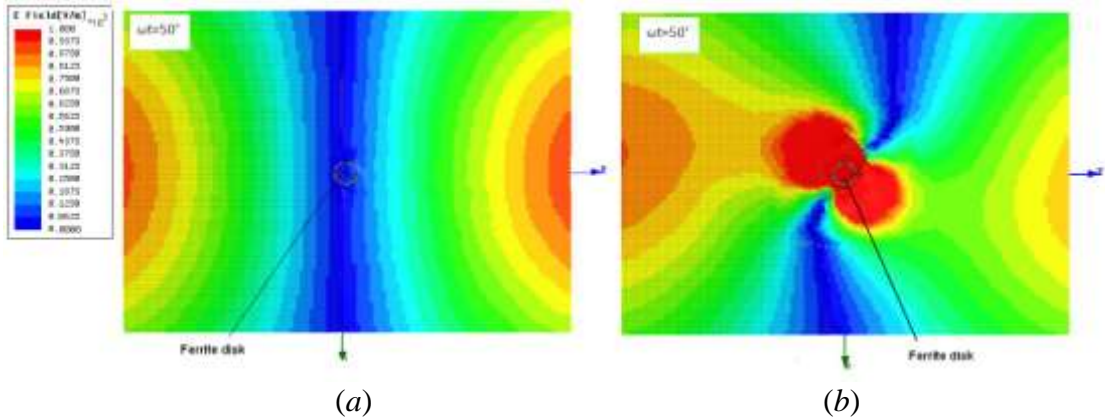

(*a*)          (*b*)

Fig. 5. The pictures of the $\vec{E}$-field structures on a vacuum plane 0.1 mm above a ferrite disk. (*a*) Undistorted wave front at frequency 3.1363GHz, which is between first and second MDM resonances; (*b*) distorted wave front at the 1st resonance frequency of 3.0237GHz.

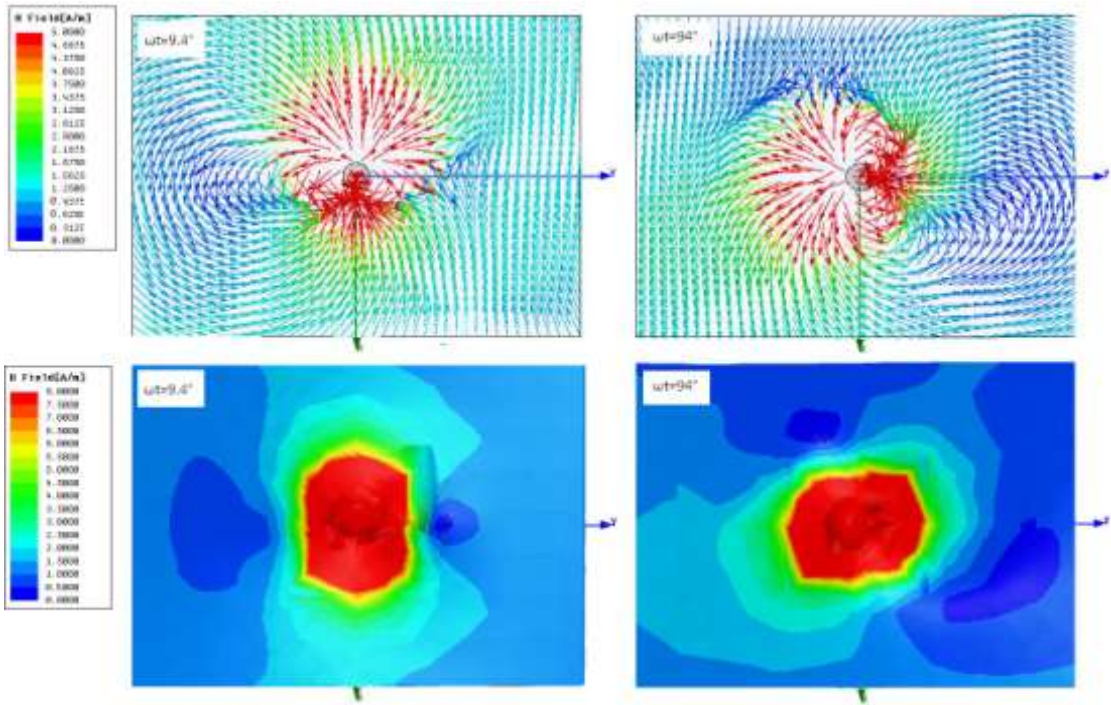

Fig. 6. Magnetic-field distributions on a vacuum plane 0.1 mm above a ferrite disk at different time phases. The frequency is the 1$^{st}$ resonance frequency of 3.0237GHz. The top row: enlarged pictures of the magnetic-field distributions immediately above a ferrite disk. The bottom row: structures of the wave fronts of the magnetic field.

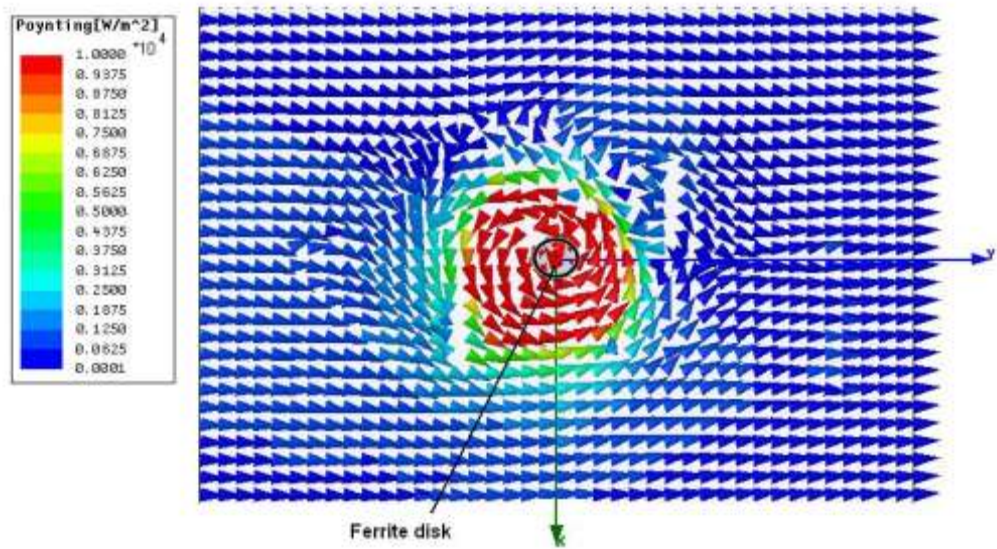

Fig. 7. Power-flow vortex on a vacuum plane 0.1 mm above a ferrite disk. The frequency is the 1$^{st}$ resonance frequency of 3.0237GHz.

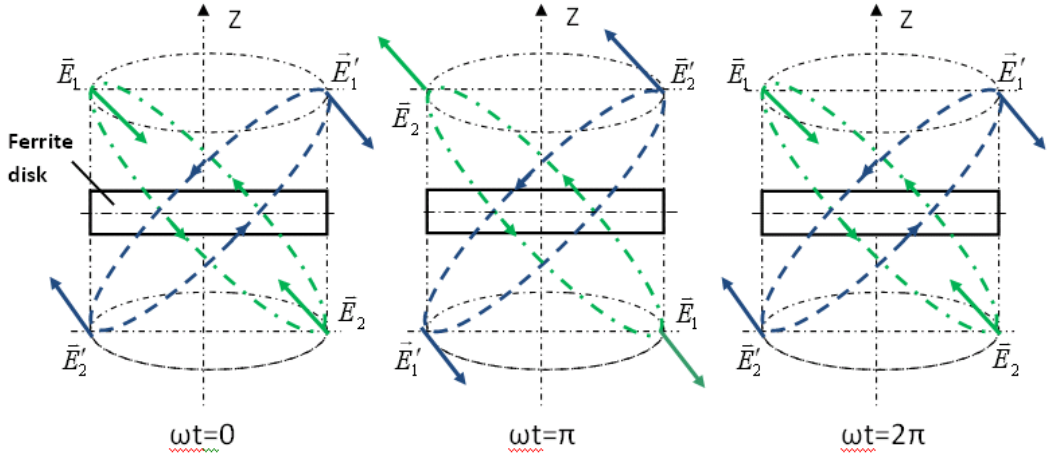

Fig. 8. Transformations of the electric fields in a region very close to a ferrite disk. The electric-field vectors are shown in connection with geometry of the MS helical-wave loops [20].

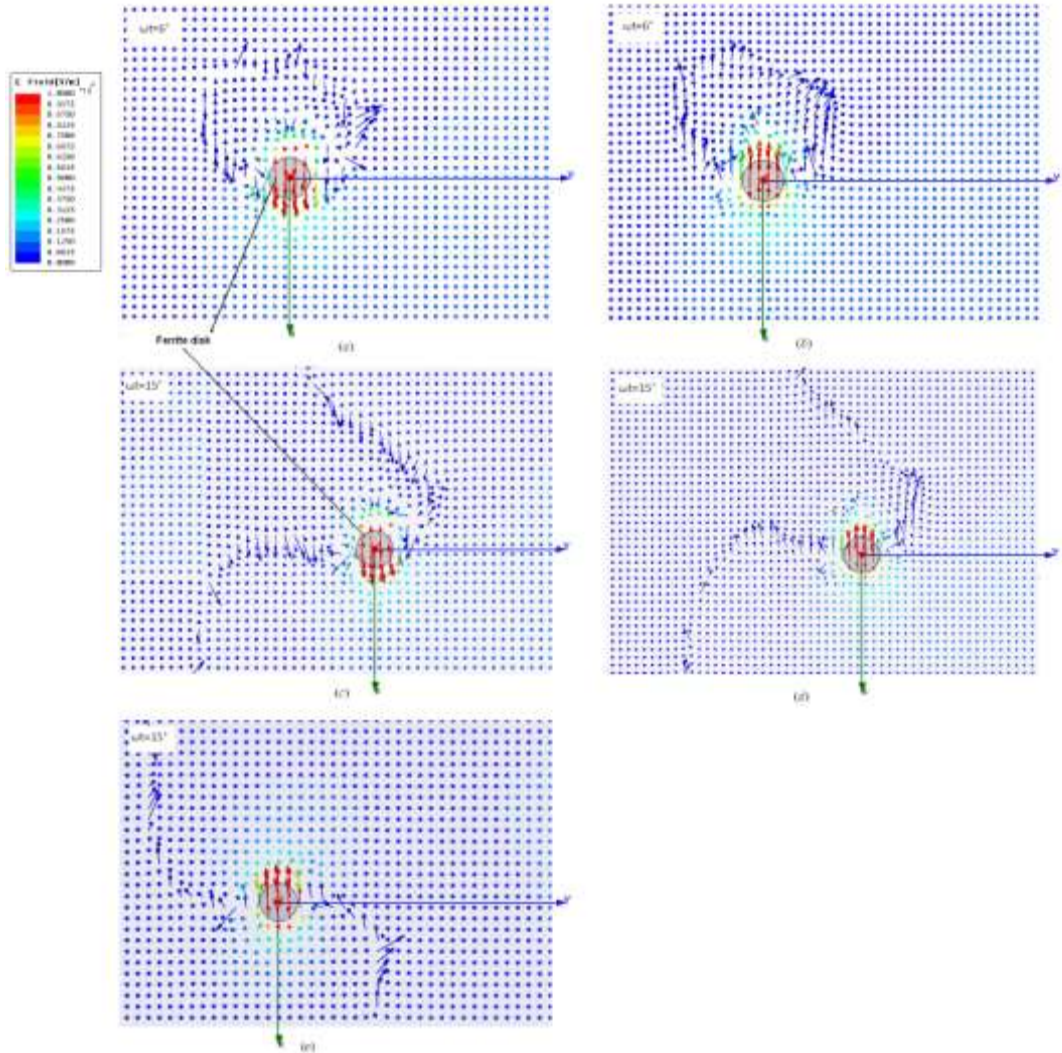

Fig. 9. The topologically originated scattered electric fields with the *xy*-plane components. (*a*), (*b*). The electric field distributions in vacuum above and below a ferrite disk, respectively, at the distances 0.1mm from the ferrite-disk planes and at a bias magnetic field directed along *z* axis. The time phase is $\omega t = 6°$. (*c*), (*d*). The electric field distributions in vacuum above and below a ferrite disk, respectively, at the distances 0.1mm from the ferrite-disk planes and at a bias magnetic field directed along *z* axis. The time phase is $\omega t = 15°$. (*e*) The same as in Fig. (d), but at a bias magnetic field directed opposite to the *z*-axis direction.

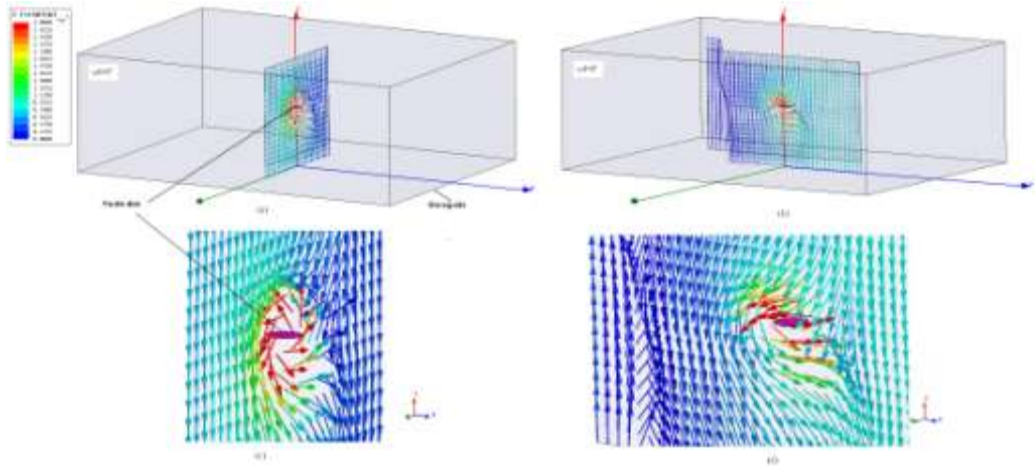

Fig. 10. The electric-field distrubution on the cross sectional planes at the 1$^{st}$ resonance frequency. (*a*) The *xz* plane in a waveguide; (*b*) The *yz* in a waveguide.

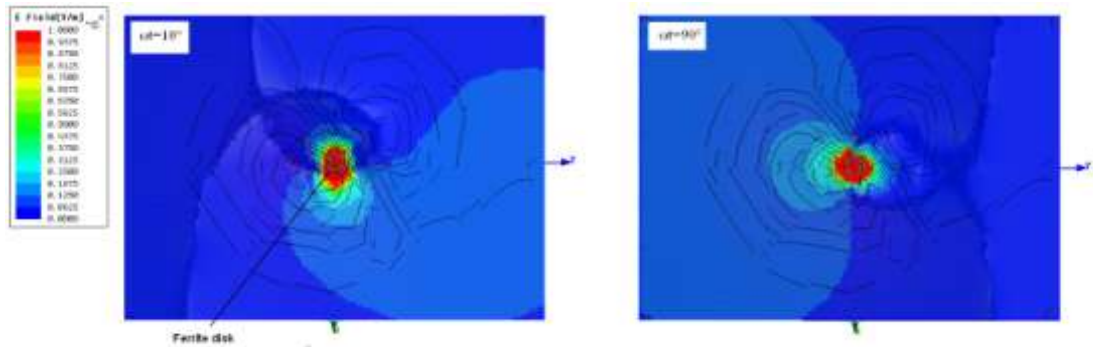

Fig. 11. The electric-field-front configuration on a vacuum plane 7 mm above a ferrite-disk plane combined with the cross-sectional outline of the helicity density on the same plane.

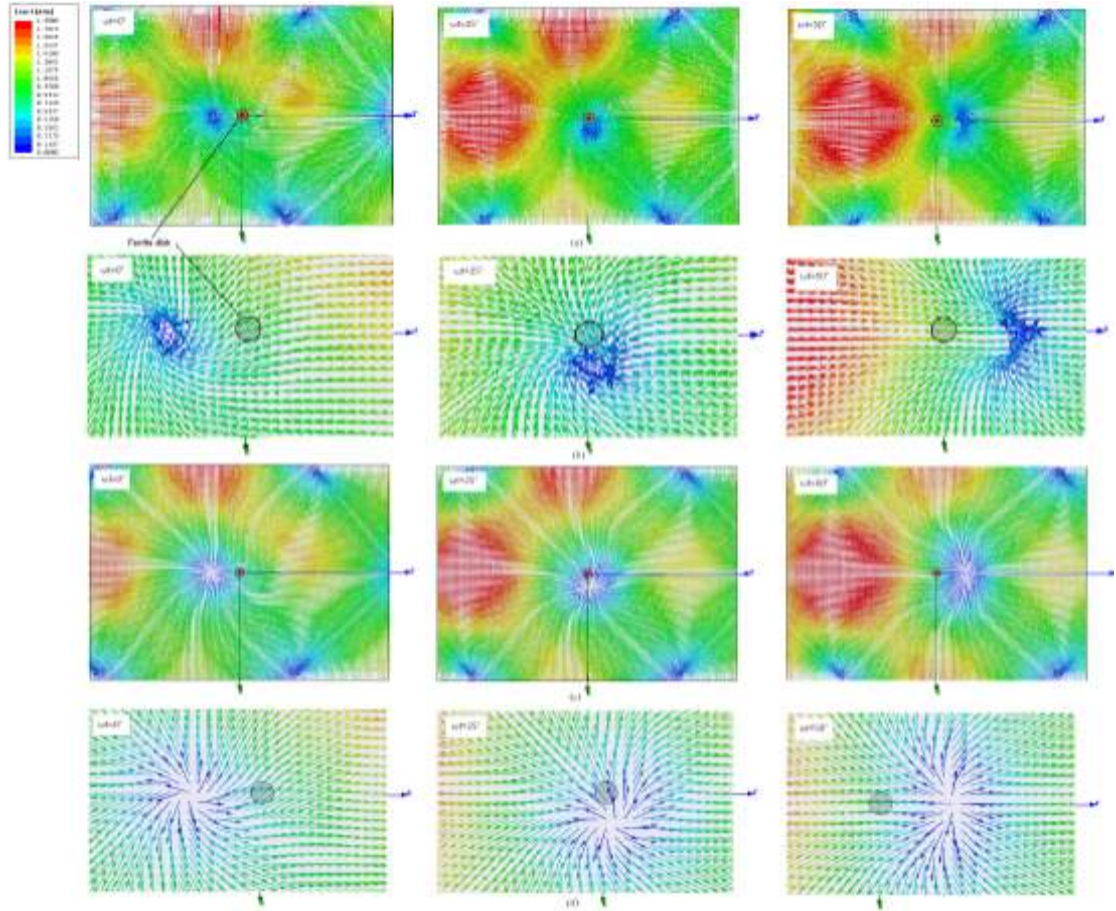

Fig. 12. Modification of the electric charge and current densities on waveguide walls at the 1st resonance frequency. Surface electric charges, appearing also as topological charges, originate surface electric currents in the forms of convergent or divergent spirals. (*a*) The electric charge and current densities on upper waveguide wall; (*b*) enlarged pictures of distributions on upper waveguide wall. (*c*) The electric charge and current densities on lower waveguide wall; (*d*) an enlarged picture of distributions on lower waveguide wall.

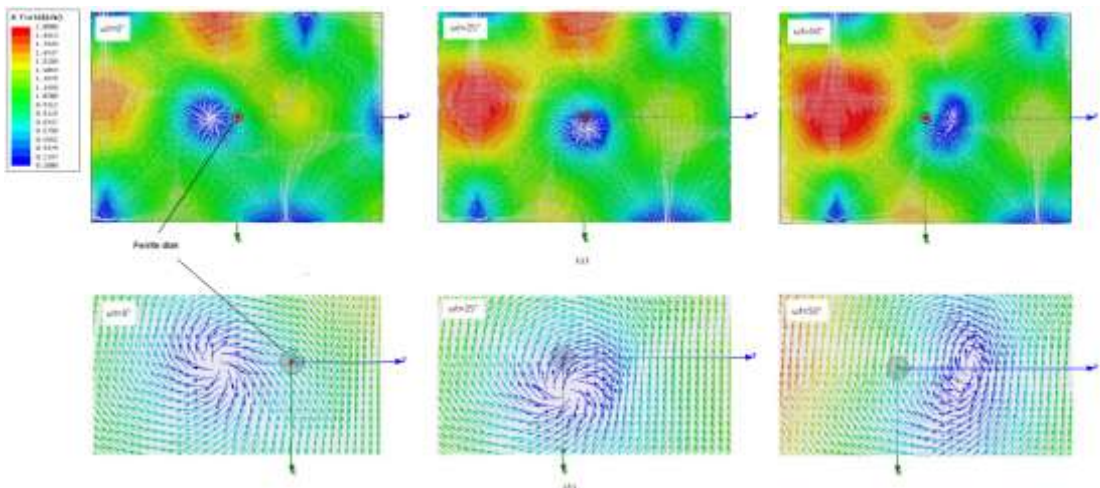

Fig. 13. Modification of the magnetic field on upper waveguide wall at the 1st resonance frequency. There is evidence for topological magnetic charges on the waveguide walls. Because of symmetry of the magnetic fields in a waveguide along *z*-axis, one has the same pictures of topological magnetic charges on the lower waveguide wall. (*a*) The magnetic field on upper waveguide wall; (*b*) enlarged pictures of distributions on upper waveguide wall.

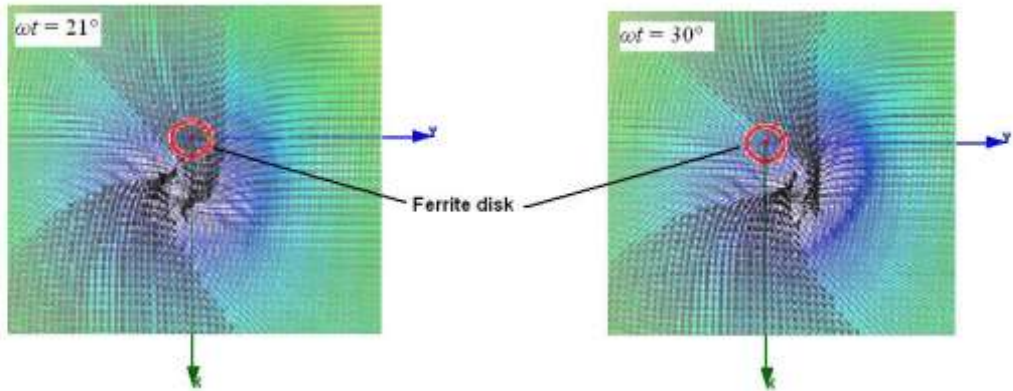

Fig. 14. A combined picture of the magnetic-field (arrows) and surface current (lines) distributions on upper waveguide wall at the 1$^{st}$ resonance frequency. (*a*) Enlarged area (with the image of a ferrite disk) for the phase $\omega t = 21°$; (*b*) the same, for the phase $\omega t = 30°$.

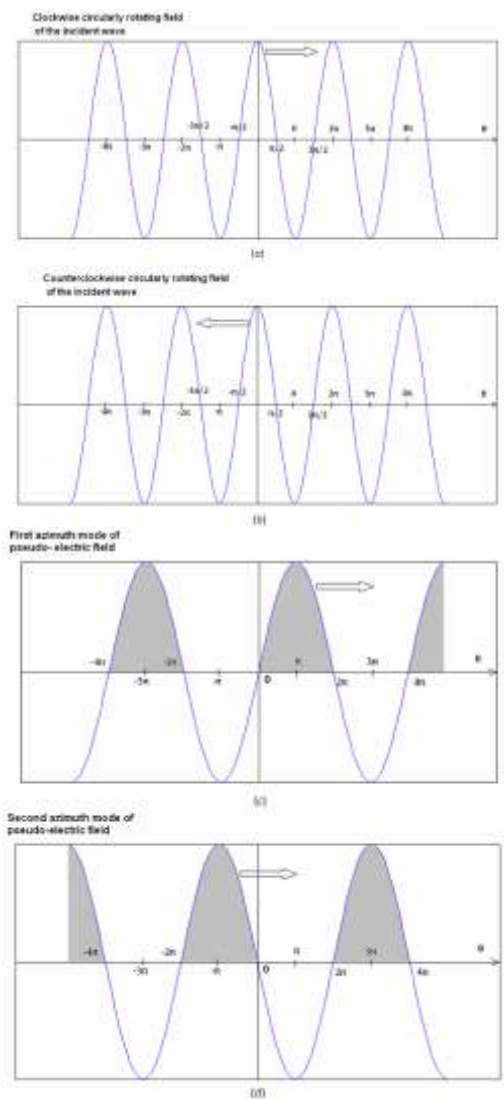

Fig. 15. The mechanism of interaction between the flux of the incident single-valued electric field and the flux of the double-valued pseudo-electric fields of MDM ferrite particles. With averaging at the azimuth coordinates, the electric flux of the incident EM field is completely annihilated by the electric flux of the MDM field. (*a*), (*b*) Two (clockwise and counterclockwise) circularly rotating $\vec{E}$-fields of the incident wave. (*c*), (*d*) Two azimuth modes of pseudo-electric fields originated from double-valued-function edge magnetic currents and rotating at a certain direction on a lateral surface of a ferrite disk.

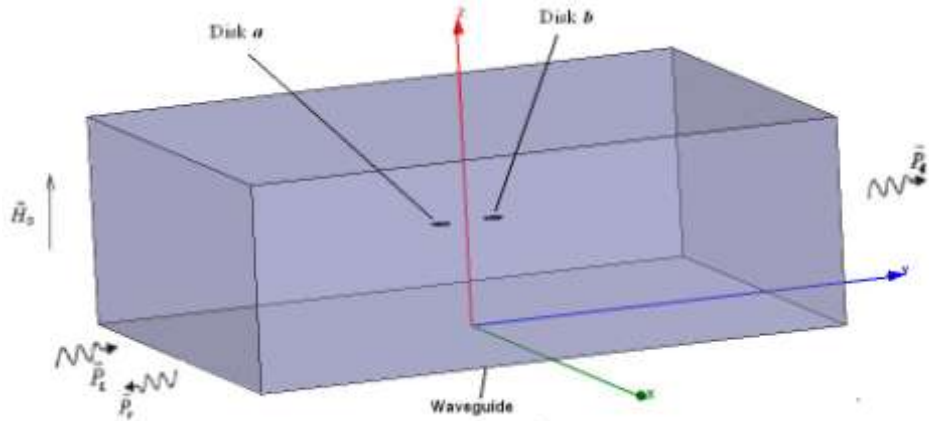

Fig. 16. A structure of two coupled MDM disks embedded in a rectangular waveguide.

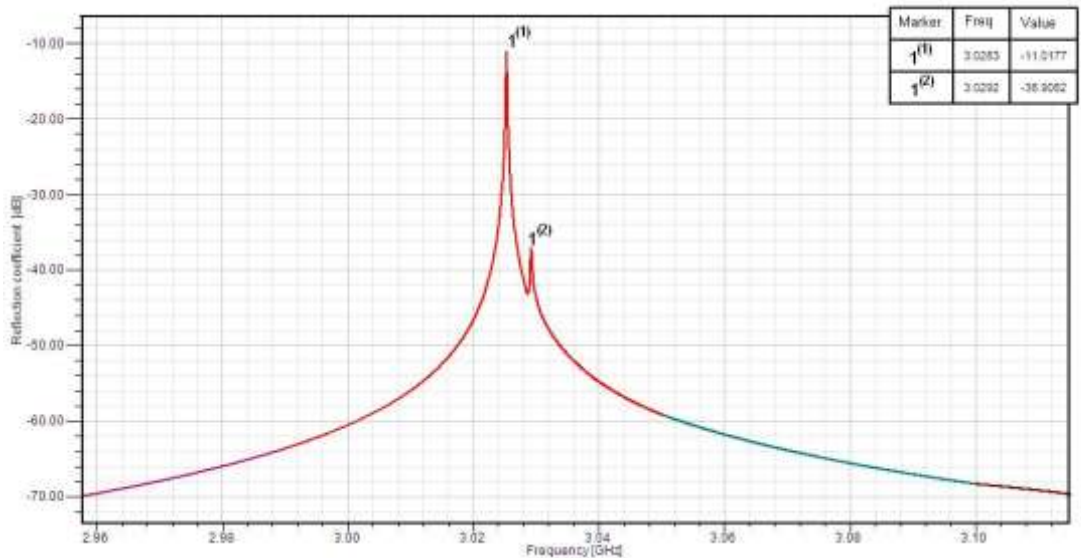

Fig. 17. Reflection coefficient for two coupled disks at the frequency region of the 1$^{st}$ MDM. The distance between centers of the disks is 3.6 mm.

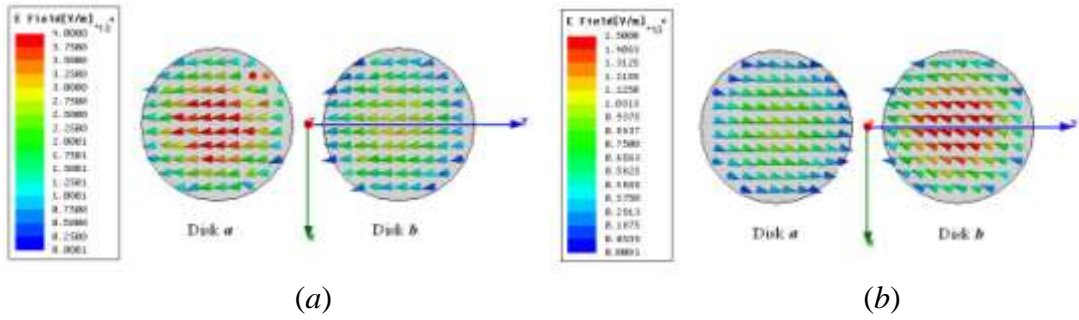

Fig. 18. The electric field distribution on the upper surface of ferrite disks. The distance between centers of the coupled disks is 3.6 mm. (*a*) At frequency (3.0253GHz) of the first peak of the splitted resonance; (*b*) at frequency (3.0292GHz) of the second peak of the splitted resonance.

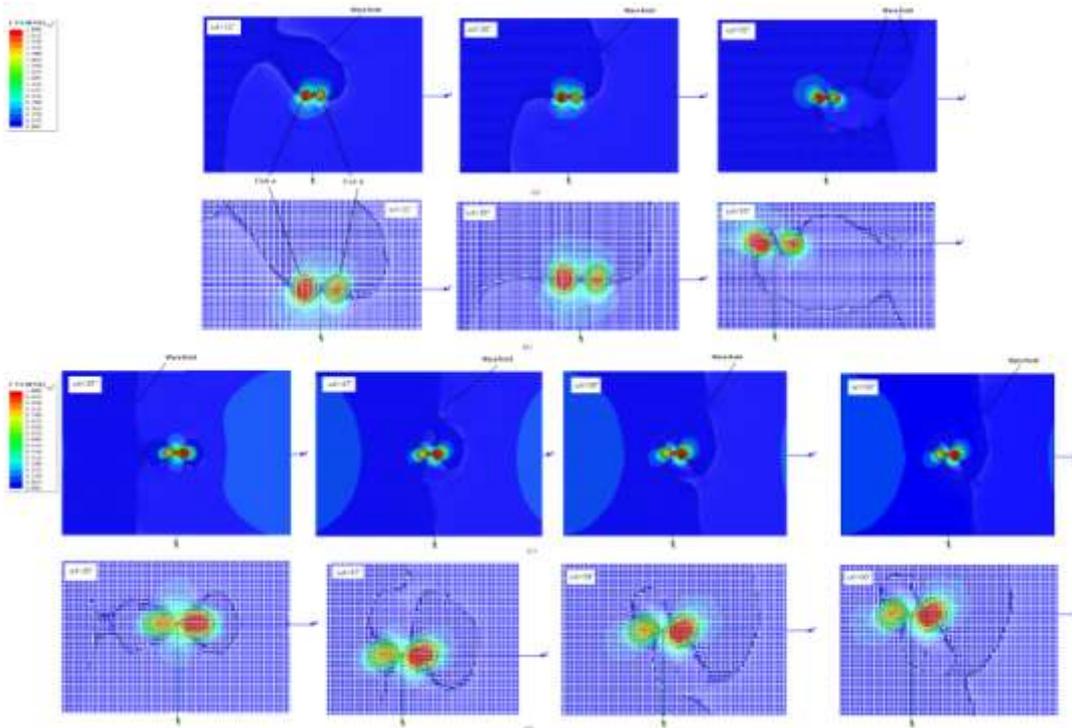

Fig. 19. Positions of the wave fronts (localized regions where the electric field of a waveguide mode changes its sign) for coupled disks on a vacuum plane passing 0.1 mm above ferrite disks. The distance between centers of the disks is 3.6 mm. (*a*) At frequency (3.0253GHz) of the first peak of the splitted resonance; (*b*) enlarged pictures of the same distributions. (*c*) At frequency (3.0292GHz) of the second peak of the splitted resonance; (*b*) enlarged pictures of the same distributions.

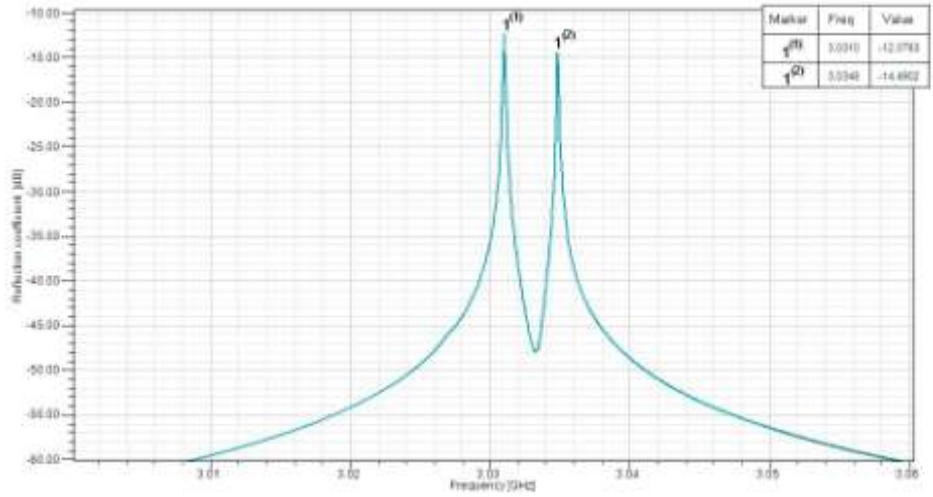

Fig. 20. Reflection coefficient for two coupled disks at the frequency region of the 1$^{st}$ MDM. The distance between centers of the disks is 6 mm.

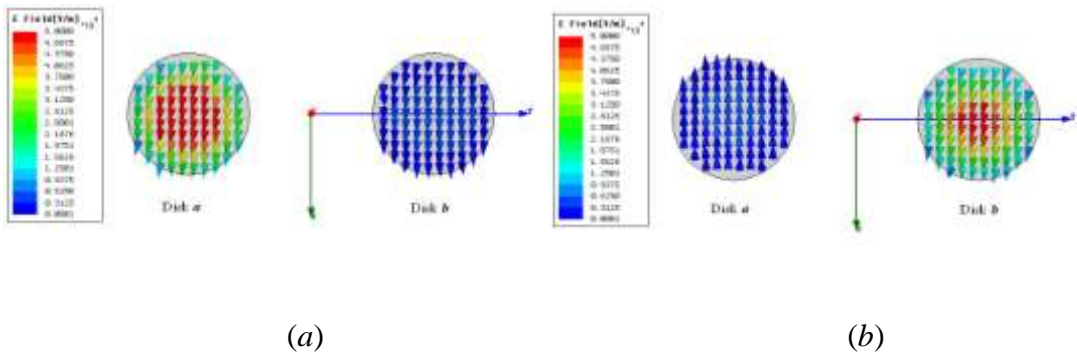

(*a*)                    (*b*)

Fig. 21. The electric field distribution on the upper surface of ferrite disks. The distance between centers of the coupled disks is 6 mm. (*a*) At frequency (3.031GHz) of the first peak of the splitted resonance; (*b*) at frequency (3.0348GHz) of the second peak of the splitted resonance.

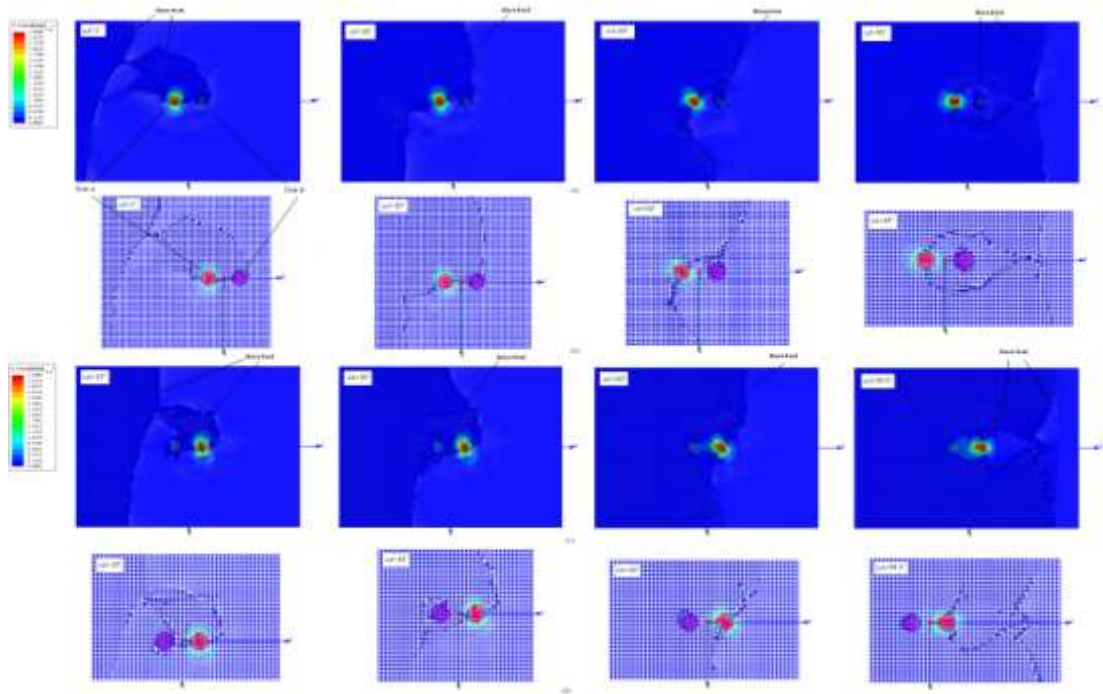

Fig. 22. Positions of the wave fronts (localized regions where the electric field of a waveguide mode changes its sign) for coupled disks on a vacuum plane passing 0.1 mm above ferrite disks. The distance between centers of the disks is 6 mm. (*a*) At frequency (3.031GHz) of the first peak of the splitted resonance; (*b*) enlarged pictures of the same distributions. (*c*) At frequency (3.0348GHz) of the second peak of the splitted resonance; (*b*) enlarged pictures of the same distributions.

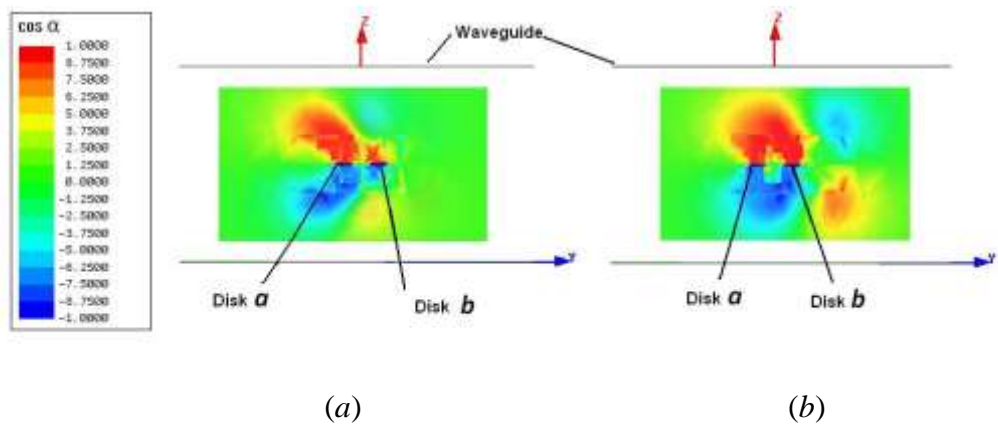

(*a*)          (*b*)

Fig. 23. The normalized helicity density for coupled disks at cross-section plane *yz* (orthogonal to a front of an incident EM wave), passing through the disk axes. The distance between centers of the disks is 6 mm. (*a*) At frequency (3.031GHz) of the first peak of the splitted resonance; (*b*) at frequency (3.0348GHz) of the second peak of the splitted resonance.

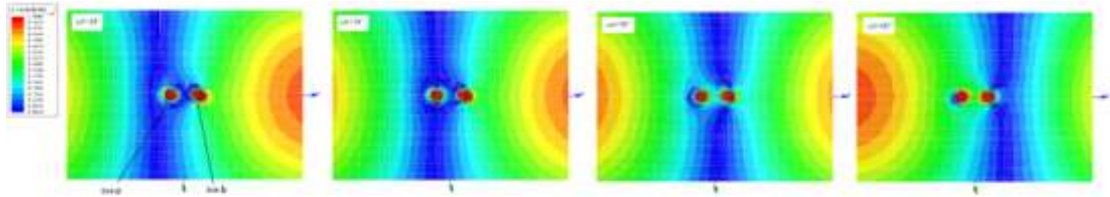

Fig. 24. The wave front of the electric field for two coupled disks at frequency 3.0333GHz between two MDM resonances. Vacuum plane is 0.1 mm above ferrite disks. The distance between centers of the disks is 6 mm.

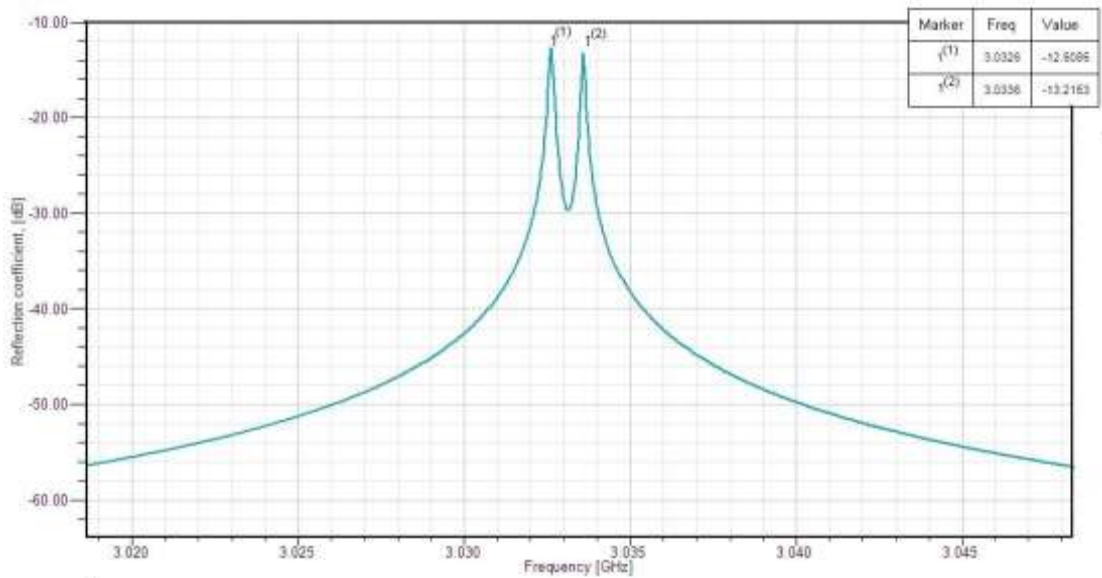

Fig. 25. Reflection coefficient for two coupled disks at the frequency region of the 1$^{st}$ MDM. The distance between centers of the disks is 12 mm.

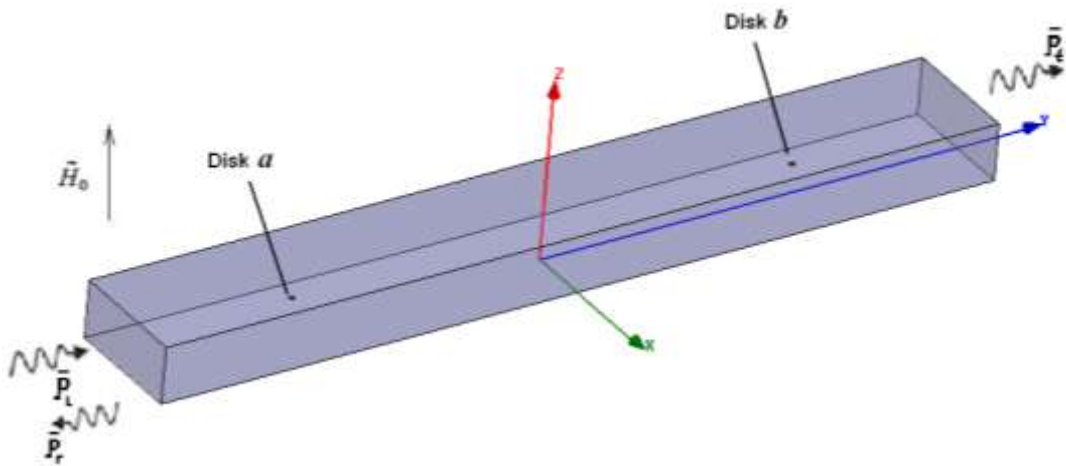

Fig. 26. A structure of two far-distant MDM disks embedded in a rectangular waveguide. The distance between the disks is 30cm.

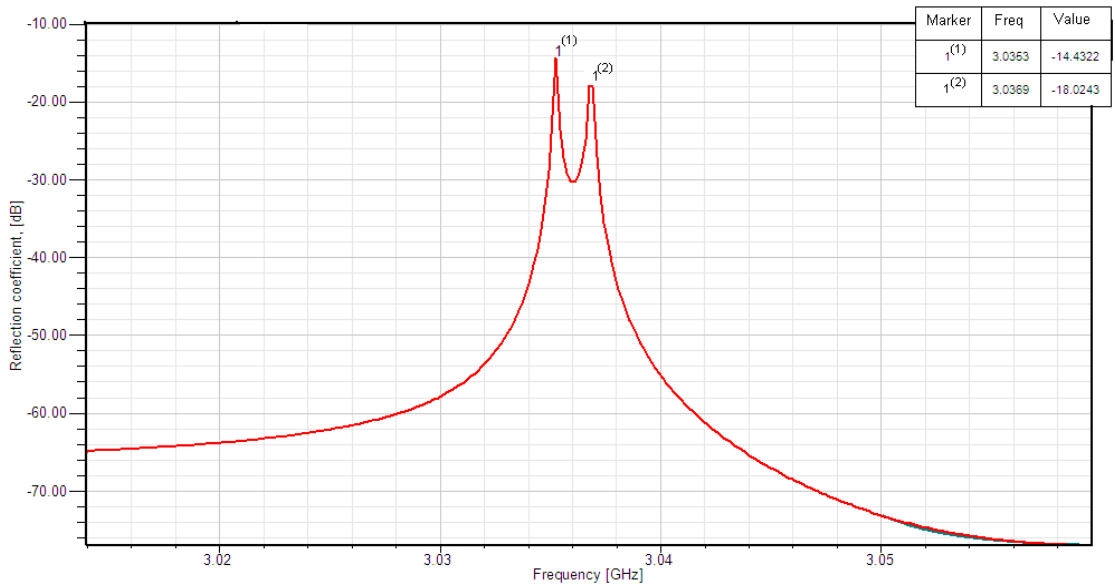

Fig. 27. Reflection coefficient for two coupled disks at the frequency region of the 1st MDM. The distance between the disks is 30cm.

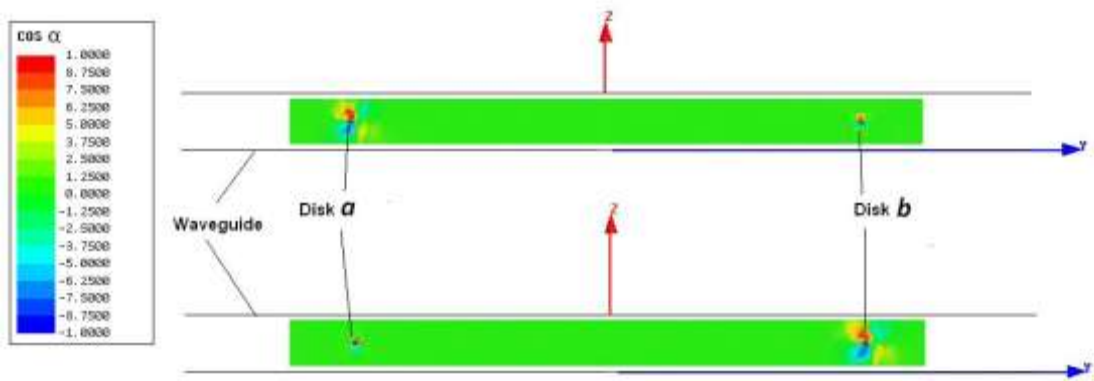

Fig. 28. The normalized helicity density for coupled disks at cross-section plane *yz* (orthogonal to a front of an incident EM wave), passing through the disk axes. The distance between the disks is 30cm. The upper picture: frequency (3.0353GHz) of the first peak of the splitted resonance. The lower picture: frequency (3.0369GHz) of the second peak of the splitted resonance.

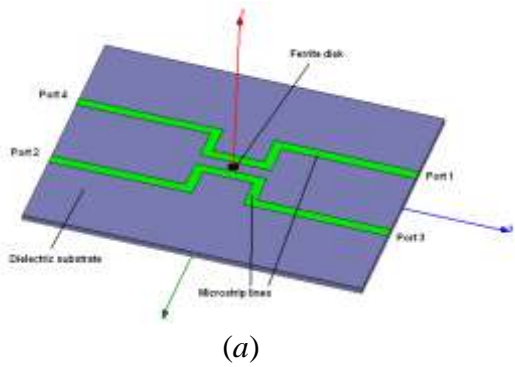　　　　　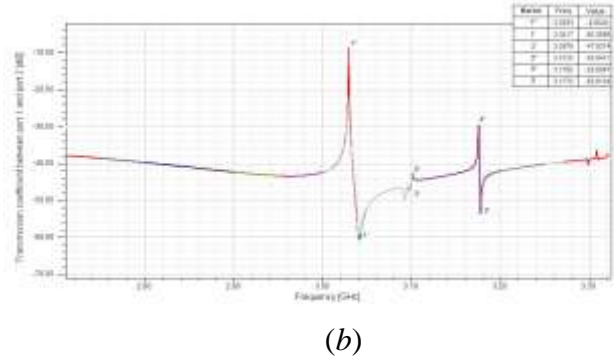

(*a*)　　　　　　　　　　　　　　　　　(*b*)

Fig. 29. A microstrip structure with an embedded ferrite disk. (*a*) Geometry of a structure (a bias magnetic field is directed along *z* axis); (*b*) a numerical spectrum for the structure (a transmission coefficient between ports 1 and 2).

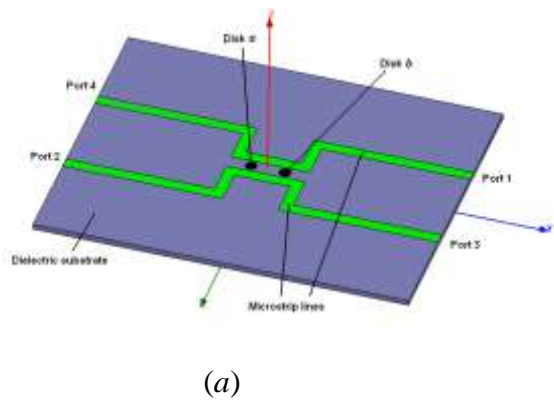　　　　　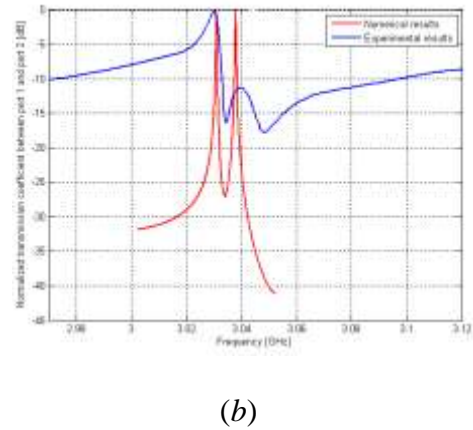

(*a*)　　　　　　　　　　　　　　　　　(*b*)

Fig. 30. A microstrip structure with two coupled ferrite disks. (*a*) Geometry of a structure (a bias magnetic field is directed along *z* axis, the distance between disks is 9mm); (*b*) numerical and experimental split-resonance characteristics for the 1$^{st}$ MDM (a normalized transmission coefficient between ports 1 and 2).